\documentclass[a4paper,11pt]{article}
\pdfoutput=1 

\usepackage{jcappub}
\usepackage[toc,page]{appendix}
\usepackage{subcaption}
\usepackage{enumerate}
\usepackage{amsmath}

\usepackage{graphicx}         
\usepackage{multirow}
\usepackage{braket}
\usepackage[T1]{fontenc} 
\usepackage[table]{xcolor}
\usepackage{float}
\usepackage{physics}
\def\noi{{\noindent}}

\newcommand{\lp}{\left(}
\newcommand{\rp}{\right)}

\def\be{\begin{equation}}
\def\ee{\end{equation}}
\def\ba{\begin{eqnarray}}
\def\ea{\end{eqnarray}}

\newcommand{\fnl}{f_{\mathrm{NL}}}

\newcommand{\ogw}{\Omega_{\mathrm{GW}}}
\newcommand{\snr}{\mathrm{SNR}}
\newcommand{\snrafg}{\mathrm{SNR}_{\text{AFG}}}

\hypersetup{pdftitle={Testing inflation on all scales: a case study with alpha-attractors}
}
\usepackage{bm}
\DeclareUnicodeCharacter{2212}{-}

\title{\boldmath Testing inflation on all scales: a case study with $\alpha$-attractors}

\author[a,b]{Laura Iacconi,}
\author[b,c,d]{Michael Bacchi,}
\author[c, d]{Luiz Filipe Guimarães}
\author[c, e]{and Felipe T. Falciano}

\affiliation{$^{a}$Astronomy Unit, Queen Mary University of London, \\
Mile End Road, London, E1 4NS, UK}
\affiliation{$^{b}$Institute of Cosmology \& Gravitation, University of Portsmouth, \\
Burnaby Road, Portsmouth, PO1 3FX, UK}
\affiliation{$^{c}$PPGCosmo, CCE, Universidade Federal do Espírito Santo (UFES), \\
Av. Fernando Ferrari, 540, CEP 29.075-910, Vitória, ES, Brazil}
\affiliation{$^{d}$Núcleo Cosmo-UFES, CCE, Universidade Federal do Espírito Santo (UFES), \\
Av. Fernando Ferrari, 540, CEP 29.075-910, Vitória, ES, Brazil}
\affiliation{$^{e}$Brazilian Center for Research in Physics (CBPF), \\ Dr. Xavier Sigaud st. 150, zip 22290-180, Rio de Janeiro, RJ, Brazil}

\emailAdd{l.iacconi@qmul.ac.uk}
\emailAdd{michael.bacchi@edu.ufes.br}
\emailAdd{luiz.f.guimaraes@ufes.br}
\emailAdd{ftovar@cbpf.br}

\abstract{
A plethora of inflationary models have been shown to produce interesting small-scale phenomenology, such as enhanced scalar fluctuations leading to primordial black hole (PBH) production and large scalar-induced gravitational waves (GW).
Nevertheless, good models must simultaneously explain current observations on all scales. 
In this work, we showcase our methodology to establish the small-scale phenomenology of inflationary models on firm grounds.
We consider the case of hybrid cosmological $\alpha$-attractors, and focus on a reduced parameter space featuring the two potential parameters which roughly determine the position of the peak in the scalar power spectrum, $\mathcal{P}_\zeta$, and its amplitude. 
We first constrain the parameter space by comparing the large-scale predictions for $\mathcal{P}_\zeta$ with current CMB anisotropies measurements and upper limits on $\mu$-distortions. 
We take into account uncertainties due to the reheating phase, and observe that the parameter-space area compatible with large-scale constraints shrinks for extended reheating stages. 
We then move to smaller scales, where we find that non-Gaussianity at peak scales is of the local type and has amplitude $f_\text{NL}\sim \mathcal{O}(0.1)$. 
This ensures that non-linear effects are subdominant, motivating us to employ the tree-level $\mathcal{P}_\zeta$ to compute the abundance of PBHs and the spectrum of induced GWs for models consistent with large-scale tests. 
The former allows us to further constrain the parameter space, by excluding models which over-produce PBHs. 
We find that a subset of viable models can lead to significant production of PBHs, and a fraction of these is within reach for LISA, having a signal-to-noise ratio larger than that of astrophysical foregrounds.
Our first-of-its-kind study systematically combines tests at different scales, and exploits the synergy between cosmological observations and theoretical consistency requirements. 
As such, it represents the first step towards establishing a paradigm for constraining inflation on all scales. 
}

\begin{document}
	\maketitle
	\flushbottom

\section{Introduction}
\label{sec: introduction}

\noi Since its first formulation, the theory of cosmological inflation has become the main paradigm for describing the early history of our cosmos, capable of solving the main issues related to the standard hot Big Bang theory, as well as explaining the origin of the large-scale structure in the universe. 
The strongest support for the inflationary paradigm resides in its agreement with the large-scale observations of the cosmic microwave background (CMB) \cite{Planck:2018jri}. 
The evidence for almost scale-invariant and approximately Gaussian fluctuations favors the simplest inflationary scenarios, single-field slow-roll (SFSR) models. 
They comprise a single, canonical scalar field, slowly rolling down its own potential, producing the background accelerated expansion and, at the same time, seeding primordial scalar perturbations through quantum-vacuum fluctuations. 

While large-scale probes point towards vanilla SFSR inflation, many inflationary models beyond SFSR allow for considerable deviations from the large-scale behavior on shorter scales, where looser constraints on the statistics of primordial perturbations apply~\cite{Gow:2020bzo}. 
This also clears the path for interesting small-scale phenomenology. 
For example, large scalar perturbations (or enhanced non-Gaussianity~\cite{Vennin:2024yzl}) could lead to the production of primordial black holes (PBHs) \cite{Carr:1974nx}, which might explain a fraction --if not the totality-- of dark matter (see, e.g., the recent review~\cite{Green:2024bam}). 
Even in the absence of PBH formation, a large peak in the power spectrum would lead to enhanced gravitational waves (GWs) sourced at second order in perturbation theory \cite{Ananda:2006af,Baumann:2007zm}.
The detection and characterization of this cosmological signal with current and future GW observatories will help us to shed light on the physics at play in the early universe, in particular on the window of observable inflation which lies beyond large-scale probes.

Throughout the years, many different mechanisms were proposed to generate a peak in the scalar power spectrum. 
Inflationary models with more than one field are particularly interesting: not only they provide a more natural framework to embed inflation in high-energy theories~\cite{Baumann:2014nda}, but also they allow for novel mechanisms, e.g. genuine multi-field behavior and geometrical effects~\cite{Palma:2020ejf, Fumagalli:2020adf, Braglia:2020eai, Iacconi:2021ltm}, which can enhance fluctuations.    
One example is given by multi-field cosmological $\alpha$-attractors, see Ref.~\cite{Achucarro:2017ing} and references therein. 
Developed in the context of supergravity, they comprise two fields living on a hyperbolic field space~\cite{Kallosh:2015zsa}. 
While single-field $\alpha$-attractors are excellent candidates to describe inflation, thanks to their universal predictions for the large-scale observables~\cite{Kallosh:2013hoa, Kallosh:2013yoa}, multi-field $\alpha$-attractors which lead to a peak in the scalar power spectrum at phenomenologically interesting scales might not be compatible with large-scale measurements of the scalar spectral tilt~\cite{Iacconi:2021ltm}. 
One possible way out consists in considering a different class of multi-field $\alpha$-attractors, namely polynomial models~\cite{Kallosh:2022feu, Braglia:2020eai, Kallosh:2022vha, Iacconi:2023slv}, which predict slightly larger values for the spectral tilt. 
Also, recently a new class of $\alpha$-attractors has been proposed~\cite{Kallosh:2022ggf,Braglia:2022phb}, developed in the context of hybrid inflation~\cite{Garcia-Bellido:1996mdl,Linde:1991km,Linde:1993cn}. 
In this scenario, the freedom to adjust the potential parameters allows one to interpolate between the usual universal prediction for $n_s$ and $n_s=1$, where a value of $n_s$ closer to unity is motivated not only in light of consistency with large-scale measurements for models with PBH production and large GW generation, but also by recent efforts to solve the $H_0$ and $S_8$ tensions~\cite{Riess:2021jrx,Abdalla:2022yfr,Jiang:2022uyg}.

As demonstrated, e.g., in Refs.~\cite{Iacconi:2021ltm, Iacconi:2023slv}, in order to establish the small-scale phenomenology of inflationary models on firm grounds, one must simultaneously take into account current observations on all scales, and theoretical consistency requirements.  
In this work we present our approach to this issue, and showcase it by studying the hybrid $\alpha$-attractors introduced in Ref.~\cite{Braglia:2022phb}. 

\subsection{Methodology}
\label{sec:methodology}
Each inflationary model is defined by its Lagrangian density, which can include many free parameters (see, e.g., Refs.~\cite{Martin:2013tda, Martin:2024qnn} for an encyclopedic classification of single-field slow-roll models). 
The Lagrangian parameters, together with initial conditions in phase space, form a multi-dimensional parameter space, in which each point represents a different realisation of the model.
Our aim is to survey the parameter space of models leading to interesting phenomenology on small scales, and single-out the volume (if any) that is consistent with current large-scale observations and theoretical requirements, and that leads to a SIGW background within reach for upcoming GW observatories. 

For each point of the parameter space, we use the publicly available \texttt{PyTransport} package~\cite{Mulryne:2016mzv, Ronayne:2017qzn} to numerically compute the primordial 2- and 3-point correlation functions, see Eqs.~\eqref{eq:def 2pt corr function} and~\eqref{eq:def 3 pt corr funct}. 
\texttt{PyTransport} provides a numerical implementation of the transport approach to primordial correlators, see e.g. Ref.~\cite{Dias:2016rjq}.
Note that enhanced fluctuations on small scales require deviations from single-field slow-roll~\cite{Motohashi:2017kbs}, and therefore it is mandatory to compute the inflationary power spectrum numerically. 
Our methodology is then articulated into four parts:
\begin{itemize}
    \item \textit{Calibrate the horizon crossing of scales.} Inflationary predictions must be connected with comoving scales. 
    We carefully calibrate the horizon crossing of scales, by taking into account the model's predictions for the duration of the observable window of inflation, and by including the effect of reheating. 
    We describe reheating as a matter-dominated phase following the end of inflation, and include its effect by studying different values of $\Delta N_\text{CMB}$, see Eq.~\eqref{eq:DeltaNCMB}, allowed over the parameter space.  
    \item \textit{Current observational constraints on large scales.} 
    We constrain the parameter space by comparing inflationary predictions with current (i) CMB measurements of the primordial power spectrum amplitude, tilt and running from \textit{Planck} 2018 and BICEP/Keck 2015 data~\cite{Planck:2018jri}; (ii) upper limit on the tensor-to-scalar ratio from \textit{Planck} 2018 and BICEP/Keck 2018 data~\cite{Planck:2018nkj,Planck:2019nip,BICEP:2021xfz, Paoletti:2022anb}; (iii) upper limit on $\mu$-type spectral distortions from COBE/FIRAS observations~\cite{Fixsen:1996nj}; (iv) CMB measurement of the amplitude of squeezed and equilateral non-Gaussianity from \textit{Planck} 2018 data~\cite{Planck:2019kim}.
    \item \textit{Theoretical consistency checks.} 
    Recently, there has been a lot of debate on the importance of 1-loop corrections in inflationary models leading to enhanced fluctuations on small scales~\cite{Cheng:2021lif, Inomata:2022yte, Kristiano:2022maq, Riotto:2023hoz, Kristiano:2023scm, Riotto:2023gpm, Firouzjahi:2023aum, Motohashi:2023syh, Firouzjahi:2023ahg, Franciolini:2023agm, Tasinato:2023ukp, Cheng:2023ikq, Fumagalli:2023hpa, Tada:2023rgp, Firouzjahi:2023bkt, Davies:2023hhn, Iacconi:2023ggt, Braglia:2024zsl, Inomata:2024lud, Ballesteros:2024zdp, Kawaguchi:2024rsv, Fumagalli:2024jzz, Green:2024fsz}. 
    Available numerical codes operate at tree-level. 
    If loop corrections are not important, it is safe to use their output to correctly evaluate the small-scale phenomenology.
    If not~\cite{Iacconi:2023slv, Fumagalli:2023loc}, more work needs to be done to obtain correct inflationary predictions\footnote{For recent developments with lattice simulations see Refs.~\cite{Caravano:2024tlp, Caravano:2024moy}.}. 
    As a first step towards a systematic check of whether perturbativity holds at peak scales, we estimate the size of 1-loop corrections to the 2-point correlation function due to cubic interactions, and compare it with the tree-level result. 
    We also implement a second theoretical test on the parameter space, by ensuring that PBHs are not overproduced.
    \item \textit{Small-scale phenomenology and future tests.}
    Large-amplitude scalar perturbations are invariably accompanied by a background of large SIGW. 
    For the parameter space regions that stand the tests described above, we compute the energy density of SIGW and from this the SNR values for future GW observatories. 
    In this work we consider the case of LISA~\cite{LISA:2017pwj, LISACosmologyWorkingGroup:2022jok} and ET~\cite{Punturo:2010zz, Maggiore:2019uih}.
    We classify as detectable the signals of models whose SNR overcomes that of expected astrophysical foregrounds~\cite{Caprini:2024ofd}.
    This allows us to single-out the volume of the parameter space (if any) that will be accessible to LISA and ET. 
\end{itemize} 
By applying our methodology, one can establish the small-scale phenomenology of inflationary models on firm grounds, and provide realistic targets for upcoming GW observatories.

\subsection{A case study with hybrid $\alpha$-attractors}
Our methodology can be applied to any inflationary model of interest, and in this work we illustrate its application by considering the case of hybrid $\alpha$-attractors~\cite{Braglia:2022phb}. 

In its original formulation \cite{Linde:1991km,Linde:1993cn}, the hybrid inflation potential is
\begin{equation}
\label{eq:original hybrid}
    V(\varphi,\, \chi) = M^2 \left[\frac{\left( \chi^2 -\chi_0^2\right)^2}{4\chi_0^2}  + \frac{\tilde m^2}{2} \varphi^2 + \frac{\tilde g^2}{2} \varphi^2 \chi^2\right] \;,
\end{equation}
where the fields $\varphi$ and $\chi$ have a canonical kinetic term. 
Here, $M^2$ is the bare mass of the $\chi$ field, and $\tilde m^2 M^2$ and $\tilde g^2 M^2$ are respectively the bare mass of the $\varphi$ field and the coupling between $\varphi$ and $\chi$.
The $\chi$ field has a mexican-hat type potential, with unstable local maximum at $\chi=0$ and global minima at $\chi=\pm \chi_0$.
The hybrid model~\eqref{eq:original hybrid} belongs to the family of exponential\footnote{One could alternatively consider hybrid polynomial $\alpha$-attractors~\cite{Braglia:2022phb}. Polynomial and exponential $\alpha$-attractors yield to different universal predictions for the large-scale observables~\cite{Kallosh:2022feu}. We leave the analysis of the parameter space of hybrid polynomial $\alpha$-attractors for future work.} $\alpha$-attractors if one of the fields, $\varphi$ in this case, is not canonical, and instead displays a kinetic term with a double pole at $\varphi^2={6\alpha}$. 
By canonically normalising the field $\varphi$ through the redefinition $\varphi \equiv \sqrt{6\alpha}\, \tanh{({\phi}/{\sqrt{6\alpha}})}$, one obtains~\cite{Braglia:2022phb}
\begin{equation} 
\label{eq:hybrid potential}
    V(\phi,\, \chi) = M^2 \left[\frac{\left( \chi^2 -\chi_0^2\right)^2}{4\chi_0^2}  + {3 \alpha \tilde m^2} \tanh^2{\left(\frac{\phi}{\sqrt{6\alpha}}\right)} + {3\alpha\tilde g^2}\tanh^2{\left(\frac{\phi}{\sqrt{6\alpha}}\right)} \chi^2 +d \chi\right] \;,
\end{equation}
where a linear term (not present in standard hybrid-inflation models) has been introduced.
This ensures that topological defects, e.g. domain walls, are not produced, while also addressing the problem of eternal inflation~\cite{Braglia:2022phb}. 
The dynamics and phenomenology of the model~\eqref{eq:hybrid potential} have been extensively studied in Ref.~\cite{Braglia:2022phb}, with emphasis on the possibility of enhancing the scalar power spectrum on small scales, and hence produce PBHs and large second-order GWs\footnote{For scenarios where the fields of the hybrid $\alpha$-attractor model~\eqref{eq:hybrid potential} are coupled with a dark fermion see Refs.~\cite{Afzal:2024hwj, Afzal:2024xci}.}.

While the parameter space of the model~\eqref{eq:hybrid potential} is multi-dimensional, for the purpose of illustrating our method it suffices to focus on a reduced slice of it. 
We vary the parameters that roughly determine the position $(\chi_0)$ and amplitude $(d)$ of the peak in the scalar power spectrum~\cite{Braglia:2022phb}. 
First, we calibrate the horizon-crossing of scales (e.g. the CMB-pivot scale), and establish which models in the parameter space are consistent with large-scale observations and constraints (Sec.~\ref{sec:large-scale tests}). 
We then move to smaller scales.
For the first time, we study the shape and amplitude of non-Gaussianity associated with the peak in the scalar power spectrum, and investigate perturbativity (Sec.~\ref{sec:Non-Gaussianities}). 
For models that are compatible with large-scale constraints, we then compute the predicted abundance of PBHs (Sec.~\ref{sec: PBH}) and the signal-to-noise ratio (SNR) of the scalar-induced GWs for LISA and ET (Sec.~\ref{sec:GW}). 
All models which (i) are compatible with large-scale constraints, (ii) do not overproduce PHBs, (iii) have SNR exceeding that of expected astrophysical foregrounds, provide well-tested and realistic targets for testing hybrid $\alpha$-attractors on small scales. 
We discuss our findings in Sec.~\ref{sec:Conclusions}.

{\textit{Notation:}}
In this work, we set $M_\text{Pl}=1$. 
Cosmic time is identified by $t$, and the Hubble rate is defined as $H\equiv \dot a /a$, where $a(t)$ is the scale factor. 
We extensively use the number of e-folds as temporal variable, defined as $N \equiv \int \mathrm{d}t \, H(t)$. 
For a general scale $k$, which crossed the horizon at time $t_k$ defined by the condition $k=a(t_k)H(t_k)$, the number of e-folds elapsed from the horizon crossing time to the end of inflation is $\Delta N_k \equiv N_\text{end} - N_k =  \int_{t_k}^{t_{\text{end}}} \mathrm{d}t\, H(t) $. 

\section{Large-scale constraints}
\label{sec:large-scale tests}

This Section is dedicated to large-scale constraints from CMB observations. 
We first briefly review the relevant features of background and perturbations evolution for models with potential~\eqref{eq:hybrid potential} (Sec.~\ref{sec: review of the model}).
Next, we focus on the reduced $(\chi_0,\,d)$ parameter space by fixing $\{\alpha=1,\, \tilde m = 0.3, \, \tilde g = 0.8\}$.
For these models\footnote{Note we use the term ``model'' to indicate both the inflationary model, e.g. Eq.~\eqref{eq:hybrid potential}, and its realisations in the parameter space.}, we discuss the horizon crossing of different scales during inflation, and the impact of an extended reheating phase (Sec.~\ref{sec: Delta N CMB}).
We then consider constraints stemming from the latest CMB measurements (Sec.~\ref{sec: CMB constraints}), and current upper limits on CMB $\mu$-spectral distortions (Sec.~\ref{sec: mu dist constraints}). 

\subsection{A brief review of background and perturbations dynamics}
\label{sec: review of the model}

The dynamics of the hybrid $\alpha$-attractor model~\eqref{eq:hybrid potential} is extensively studied in Ref.~\cite{Braglia:2022phb}. 
Here we briefly review the most notable features of the background and scalar perturbation evolution.

First, we note that the potential is invariant under $\chi_0\to -\chi_0$, so 
for simplicity in the following we will assume $\chi_0>0$. 
We also assume that for arbitrarily large values of $\phi$, the squared mass of $\chi$ at $\chi=0$,
\begin{equation}
    \label{eq:squared mass chi}
    m_\chi^2 \equiv M^2 \left(-1+ 6 \alpha \tilde g^2 \tanh^2{\left(\frac{\phi}{\sqrt{6\alpha}}\right)}  \right)\;, 
\end{equation}
is positive, which is ensured by the condition $6\alpha \tilde g^2>1$. 
As discussed in Sec.~\ref{sec: introduction}, the hybrid model~\eqref{eq:hybrid potential} is accommodated within the family of $\alpha$-attractors. 
Therefore, $\alpha$ should be interpreted as a positive parameter related to the underlying geometrical formulation of $\alpha$-attractors within supergravity~\cite{Carrasco:2015uma, Kallosh:2015zsa}. 

For finite $\phi$, the sign of the squared mass~\eqref{eq:squared mass chi} depends on whether $\phi$ is larger or smaller than a critical value, $\phi_\text{c}$, defined by $\tanh^2{(\phi_\text{c}/\sqrt{6\alpha})} \equiv  (6\alpha \tilde g^2)^{-1}$. 
For $\phi\gg \phi_\text{c}$, the squared mass of $\chi$ is positive, stabilising $\chi$ at the local minimum 
\begin{equation}
\label{eq:chi valley}
    \chi_\text{valley}=\frac{-d}{6\alpha \tilde g^2 \tanh^2{\left(\frac{\phi}{\sqrt{6\alpha}}\right)} -1}  \;,
\end{equation}
which is slightly ($|d|\ll 1$) displaced from $\chi=0$. 
For trajectories starting from $\phi=\phi_\text{in}>\phi_\text{c}$ and $\chi=\chi_\text{in}$, the field rapidly evolves towards $\chi=\chi_\text{valley}$
and then proceeds in the $\phi$ direction.
Note that as long as $\phi_\text{in}\gg \chi_\text{in}$ the field $\chi$ is stabilised at $\chi_\text{valley}$ well before the CMB scale crossed the horizon. 
This makes the background evolution during the observable window of inflation independent of the initial conditions~\cite{Braglia:2022phb}.
Along the $\chi=\chi_\text{valley}$ minimum, the potential for $\phi$ is that of an $\alpha$-attractor T-model~\cite{Kallosh:2013hoa}, uplifted by a constant term, $V_\text{up}=M^2 \chi_0^2/4$.

The introduction of a linear term in Eq.~\eqref{eq:hybrid potential} ensures that topological defects, e.g. domain walls, are not produced, while also addressing the problem of eternal inflation~\cite{Braglia:2022phb}.
Without loss of generality\footnote{Note that we could equally assume that $d>0$, in which case $\chi_\text{valley}<0$ and, for $\phi<\phi_\text{c}$, the inflaton rolls towards the minimum at $\chi_\text{min} = -\chi_0 -d/2+ \mathcal{O}(d^2)$. Also in this case, one has $V(\phi_\text{min}, \chi_\text{min}) = - M^2 d \chi_0 + \mathcal{O}(d^2)<0$, which can be offset by introducing a constant, positive shift in the potential~\eqref{eq:hybrid potential}, see footnote~\ref{footnote:negative V at the minimum}.}
we assume $d<0$, in which case $\chi_\text{valley}$ is a small, positive number.
When $\phi<\phi_\text{c}$, the $\chi$ field becomes tachyonic, i.e. its squared mass~\eqref{eq:squared mass chi} turns negative, and evolves away from $\chi_\text{valley}$, towards the global potential minimum at $\phi=0$ and $\chi = \chi_\text{min}$.
For $\chi_0\ll 1$, $\chi$ is very heavy ($m_\chi^2\gg H^2$) and inflation ends abruptly. 
On the other hand, for larger values of $\chi_0$, $\chi$ is allowed to slow roll and drive  a second phase of inflation.
This is the regime Ref.~\cite{Braglia:2022phb} investigates, and which will also be our focus.

Due to the fact that $\chi_\text{valley}>0$, the field rolls towards larger $\chi$ values
, with the minimum  $\chi_\text{min}$ slightly displaced from $\chi_0$ due to $d\neq 0$.  
By measuring the displacement in terms of powers of $d$, we find 
\begin{equation}
    \label{eq:chi min}
    \chi_\text{min} = \chi_0  - \frac{d}{2} + \mathcal{O}(d^2)\;,
\end{equation}
where higher-order coefficents can be found by solving $V_\chi|_{\chi_\text{min}}=0$ order by order\footnote{
    \label{footnote:negative V at the minimum}
    At the minimum, the potential is given by the negative, small value $V(\phi_\text{min}, \chi_\text{min}) = M^2 d \chi_0 + \mathcal{O}(d^2)$. 
    In order to have vanishing potential energy at the minimum, we consider to add a constant, positive contribution to the potential~\eqref{eq:hybrid potential}, $\Delta V = M^2 |d| \chi_0$. 
    Since this term would have no consequences for the inflationary dynamics, we have not explicitly included it in Eq.~\eqref{eq:hybrid potential}.}.

\begin{figure}
\centering
\captionsetup[subfigure]{justification=centering}
\begin{subfigure}[b]{0.49\textwidth}
\includegraphics[width=\textwidth]{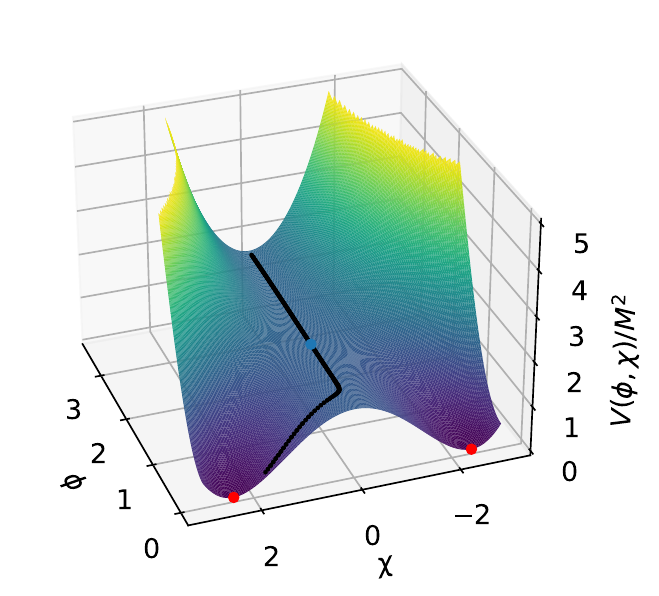}
\end{subfigure}
\begin{subfigure}[b]{0.49\textwidth}
\includegraphics[width=\textwidth]{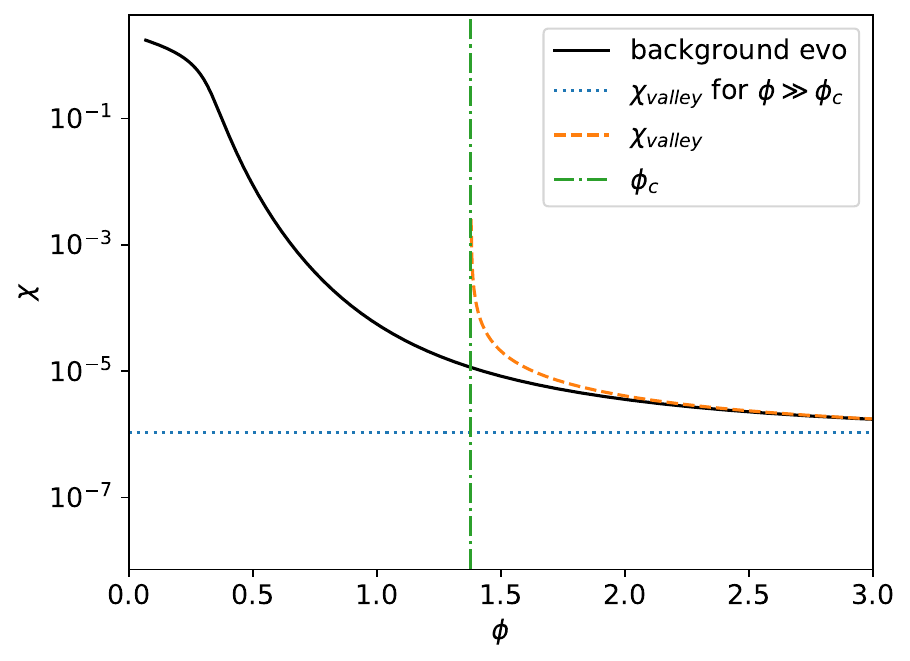}
\end{subfigure}
\caption{\textit{Left panel:} background trajectory (black line) corresponding to the parameters $\{\chi_0=2.4, \, d=-3\times 10^{-6}, \, \alpha=1,\, \tilde m = 0.3, \, \tilde g = 0.8\}$, represented on top of the potential profile, see Eq.~\eqref{eq:hybrid potential}. The blue point identifies the point along the trajectory when $\phi=\phi_\text{c}$, and the red points correspond to the minima of the potential, with the one on the left being the global minimum, see Eq.~\eqref{eq:chi min}. \textit{Right panel:} background trajectory in the $(\phi,\,\chi)$ plane (black, continuous line) for the same model represented in the left panel. The coloured lines show respectively the value of $\chi_\text{valley}$ as a function of $\phi$ (dashed, orange line), its limiting, constant value for $\phi\gg \phi_\text{c}$ (dotted, blue line) and $\phi=\phi_\text{c}$ (dot-dashed, green line).}
\label{fig:background evo}
\end{figure}
In order to illustrate the background dynamics discussed above, we display in Fig.~\ref{fig:background evo} an explicit example of it, numerically obtained with \texttt{PyTransport}~\cite{Mulryne:2016mzv, Ronayne:2017qzn, Dias:2016rjq} for a fixed potential shape (see the caption for more details).  
In the left panel one can see the fields evolution represented on top of the potential profile. 
Due to the smallness of the $d$ parameter, it is impossible to visualise by eye the displacement of $\chi_\text{valley}$ from $\chi=0$ in the left panel. 
This is displayed explicitly in the right panel, where the same background evolution is plotted in field space. 

The statistics of quantum scalar perturbations produced during inflation is determined by the correlation functions $\langle \hat\zeta(\mathbf{x}_1) \,\hat\zeta(\mathbf{x}_2) \cdots \hat\zeta(\mathbf{x}_n)\rangle$, where $\zeta$ is the comoving curvature perturbation and the hat symbol is a reminder of the fact that $\hat \zeta$ is a quantum operator. 
In the following we will drop the hat symbol for simplicity. 
The 2-point correlator, $\langle \zeta(\mathbf{x}_1) \,\zeta(\mathbf{x}_2)\rangle$, is related to the variance of $\zeta$, and from its Fourier transform one can define the dimensionless primordial scalar power spectrum, $\mathcal{P}_\zeta(k)$, as
\begin{equation}
    \label{eq:def 2pt corr function}
    \langle \zeta(\mathbf{k}_1) \zeta(\mathbf{k}_2)\rangle \equiv  (2\pi)^3 \delta^3(\mathbf{k}_1 + \mathbf{k}_2) \frac{2\pi^2}{k^3} \mathcal{P}_\zeta(k)\;. 
\end{equation}
For Gaussian perturbations, the statistics of $\zeta$ is fully determined by $\mathcal{P}_\zeta(k)$.
If $\zeta$ is a non-Gaussian variable, the first quantity which describes deviation from Gaussianity is the 3-point correlation function, whose Fourier transform is the primordial bispectrum. 
We will discuss non-Gaussianity in Sec.~\ref{sec:Non-Gaussianities}, while for the moment we focus on $\mathcal{P}_\zeta(k)$. 

Whilst CMB measurements point to small, and almost Gaussian perturbations on large scales ($k_\text{CMB}=0.05\, \text{Mpc}^{-1}$), 
Ref.~\cite{Braglia:2022phb} showed that the model~\eqref{eq:hybrid potential} can lead to large perturbations on small scales, which correspond to a peak in the primordial power spectrum at $k\gg k_\text{CMB}$. 
This is produced by the interplay between curvature and isocurvature perturbations. 
When $\phi<\phi_\text{c}$, the isocurvature perturbations become tachyonic and transiently grow on super-horizon scales. 
Once the $\chi$ field starts rolling towards its minimum, the resulting turn in field space allows the enhanced isocurvature modes to source the curvature perturbation, yielding a peak in $\mathcal{P}_\zeta(k)$. 

While all the potential parameters conspire to fix the shape of $\mathcal{P}_\zeta(k)$, in Ref.~\cite{Braglia:2022phb} it is shown that the peak position is roughly determined by $\chi_0$, which sets the duration of the second phase of evolution, driven by $\chi$, and the peak amplitude is mostly regulated by $d$, which has to be fine-tuned in order to have sufficient (but not excessive, see Sec.~\ref{sec: PBH}) amplification of perturbations on small scales. 
Therefore, we will focus on the reduced parameter space $(\chi_0, \, d)$, and fix $\{\alpha=1,\, \tilde m = 0.3, \, \tilde g = 0.8\}$. 
Note that this choice is in accordance with our goal.
We do not aim at comprehensively testing the multi-dimensional parameter space of hybrid $\alpha$-attractors.
Instead, we want to illustrate our method to establish the small-scale phenomenology of inflationary models by combining tests and observations on all scales. 

\subsection{Calibrating the horizon-crossing of scales}
\label{sec: Delta N CMB}
In order to compute large-scale predictions of a specific inflationary model (as well as those on small scales), one needs to connect scales at which different experiments constrain primordial observables to the times during inflation when said scales left the horizon. 
The key quantity calibrating this connection is the duration of inflation (measured in e-folds) since the CMB scale crossed the horizon, $\Delta N_\text{CMB}$.
We fix the CMB pivot scale to be $k_\text{CMB}=0.05\,\text{Mpc}^{-1}$, as it is best constrained from CMB measurements~\cite{Planck:2018jri}.
Following~\cite{Planck:2018jri}, one has
\begin{equation}
\label{eq:DeltaNCMB}
    \Delta N_{\text{CMB}} \equiv N_{\text{end}} - N_{\text{CMB}} \simeq  61.02 + \frac{1}{4}\ln{\left(\frac{V^2_{\text{CMB}}}{\rho_{\text{end}}}\right)} - \frac{1-3w}{4}\Delta \tilde{N}_{\text{rh}} \;.
\end{equation}
Here, $\rho_{\text{end}}$ is the energy density at the end of inflation, $V_{\text{CMB}}$ is the value of the potential at the horizon crossing of the CMB scale, and $w$ and $\Delta \tilde{N}_{\text{rh}}$ are parameters describing the reheating process. 
Note that while one normally expects non-perturbative  mechanisms to play a role during reheating (see, e.g., the lecture notes~\cite{Lozanov:2019jxc}), modelling the details of reheating is beyond the scope of this work.
For this reason, we consider perturbative reheating.

The parameter $w$ in Eq.~\eqref{eq:DeltaNCMB} represents the effective equation-of-state parameter during reheating, whose duration is given by 
\begin{equation}
\label{eq:max reheating duration}
    \Delta \tilde{N}_{\text{rh}} \equiv N_{\text{rh}} - N_{\text{end}} = \frac{1}{3(1+w)} \ln \left(\frac{\rho_{\text{end}}}{\rho_{\text{th}}}\right) \;, 
\end{equation}
where $\rho_{\text{th}}$ is the energy scale at the end of reheating. 
If the effective equation-of-state parameter for reheating is $-1<w<1/3$, $\Delta N_{\text{CMB}}$ is maximized in the instant reheating scenario ($\rho_{\text{end}}=\rho_{\text{th}}$, or equivalently $\Delta \Tilde{N}_{\text{rh}}=0$).
In the following, we assume matter-dominated reheating\footnote{In the vicinity of the minimum $(\phi = 0, \, \chi\approx \chi_0)$, the potential~\eqref{eq:hybrid potential} can be approximated as $V(\phi, \, \chi) \approx M^2 \left[\left( \chi-\chi_0\right)^2  +\frac{\tilde m^2 \phi^2}{2}  +\frac{ \tilde g^2}{2} \phi^2 \chi^2 \right]$, where we have included up to quadratic terms.}, and therefore set $w=0$.
As reheating needs to be completed at sufficiently high temperatures to allow for baryogenesis after inflation, which itself requires physics beyond the Standard Model~\cite{Mukhanov:2005sc}, we choose $\rho_\text{th}\geq (1 \, \text{TeV})^4$, yielding a (very conservative) upper limit on the duration of reheating
\begin{equation}
    \Delta \tilde{N}_{\text{rh}} \leq \Delta \tilde{N}_{\text{rh,max}} = \frac{1}{3} \log \frac{\rho_{\text{end}}}{(1 \, \text{TeV})^4}
    \;.
\end{equation}
 
As a first step towards a realistic estimation of $\Delta N_\text{CMB}$, we compute its value in the case of instant reheating, $\Delta N_{\text{CMB, inst rh}}$. 
By fixing the potential parameter $M^2$, see Eq.~\eqref{eq:hybrid potential}, such that the primordial power spectrum has the correct amplitude at CMB scales\footnote{\label{footnote: Msq}To build the numerical function $M^2(\Delta N_\text{CMB, inst rh})$ we employ \texttt{PyTransport}. For each model in the parameter space, we define this as an interpolating function built from 20 (numerically found) points with $\Delta N_\text{CMB, inst rh}\in [50,60]$. These points are obtained by requiring that $\mathcal{P}_\zeta(k_\text{CMB})$ is within $0.1\%$ of its measured central value, $\ln{\left(10^{10} A_s \right)} =  3.044 \pm 0.014$ ($68\%$ C.L.)~\cite{Planck:2018jri}. We test the $M^2(\Delta N_\text{CMB, inst rh})$ function by computing $\mathcal{P}_\zeta(k_\text{CMB})$ for 30 randomly-generated points in the interval above, and checking that the values returned are always within the $95\%$ C.L. interval. As expected, for models which display multi-field behavior when the CMB scale crossed the horizon, the numerically-found function $M^2(\Delta N_\text{CMB, inst rh})$ strongly deviates from the values one would na\"{i}vely obtain by using the single-field slow-roll expression for $\mathcal{P}_\zeta$.}, we compute $\Delta N_{\text{CMB, inst rh}}$ by iteratively solving Eq.~\eqref{eq:DeltaNCMB}. 
\begin{figure}
\begin{center}
\includegraphics[width=.7\linewidth]{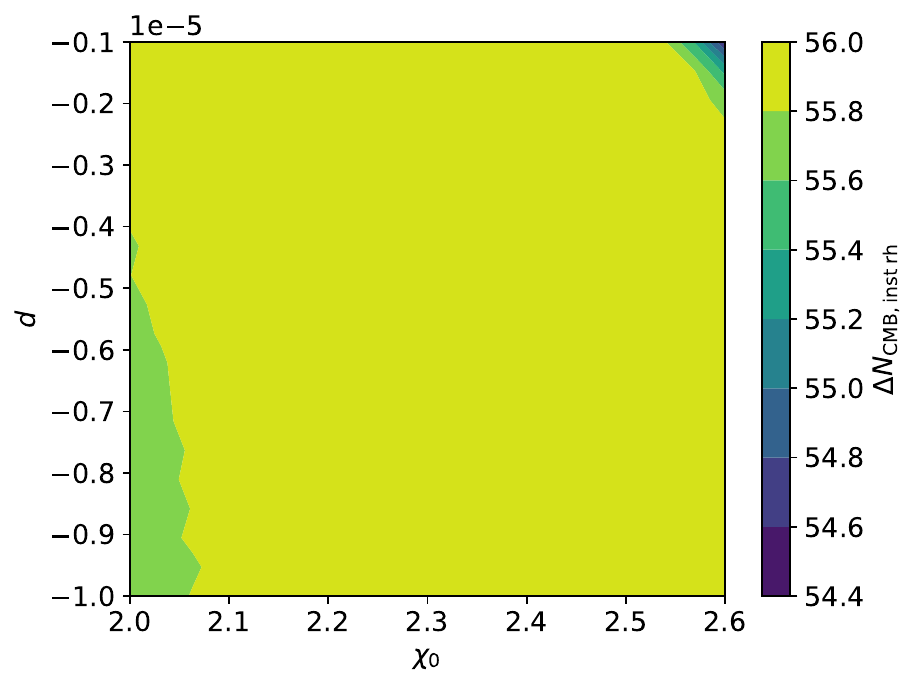}
\caption{Numerically produced values for the number of e-folds elapsed from the horizon crossing of the CMB scale to the end of inflation, see Eq.~\eqref{eq:DeltaNCMB}, obtained by assuming instant reheating. The models considered have inflationary potential~\eqref{eq:hybrid potential}, with fixed $\{\alpha=1,\, \tilde m = 0.3, \, \tilde g = 0.8\}$.}
\label{fig:InstReh_Chi0D}
\end{center}
\end{figure}
The results are shown in Fig.~\ref{fig:InstReh_Chi0D}, where one sees that there is a small variation of $\Delta N_\text{CMB, inst rh}$ over the parameter space, $ 54.597 \leq \Delta N_\text{CMB, inst rh} \leq 55.901$, with the smallest value corresponding to the model $\{\chi_0 =2.6 ,\, d=-10^{-6}\}$ and the largest to $\{\chi_0 =2.474 ,\, d=-10^{-6}\}$.
The rapid variation of $\Delta N_\text{CMB, inst rh}$ in the top-right corner of the parameter space is due to the fact that for those models multi-field effects are already relevant at CMB scales and the scale of inflation is lower as a consequence (see footnote~\ref{footnote: Msq}). 

Starting from the results for $\Delta N_\text{CMB, inst rh}$, we find the allowed values of $\Delta N_\text{CMB}$ for each model in the parameter space by including the effect of reheating as in Eq.~\eqref{eq:DeltaNCMB}. 
In particular, for each model $\Delta N_\text{CMB, min} \leq \Delta N_\text{CMB} \leq \Delta N_\text{CMB, inst rh}$, where $\Delta N_\text{CMB, min}$ corresponds to the case reheating lasted $\Delta \tilde N_\text{rh, max}$ e-folds\footnote{We numerically compute the maximum duration of reheating using Eq.~\eqref{eq:max reheating duration}. We obtain $ 37.86 \leq \Delta \tilde N_\text{rh, max}\leq 39.62$ over the whole parameter space, with smaller values in the top-right corner. This is again a consequence of having smaller $H$ in those models (see footnote~\ref{footnote: Msq}), which in turns lowers $\rho_\text{end}$.}.   
From the allowed range of $\Delta N_\text{CMB}$ computed for each point in the parameter space, we select three values which are compatible with all models and which will be the focus of our analysis for this work: $\Delta N_\text{CMB}=\{50, 52.5, 54\}$. 
The three selected values correspond to different duration of reheating over the parameter space, which for each model is given by $\Delta \tilde N_\text{rh} = 4 \left( \Delta N_\text{CMB, inst rh} - \Delta N_\text{CMB}\right)$, see Eq.~\eqref{eq:DeltaNCMB}. 
Specifically,
\begin{equation}
    \begin{aligned}
    \label{eq:reheating duration benchmark}
    \Delta N_\text{CMB} = 50 & \;\longrightarrow\;  18.39\leq \Delta \tilde N_\text{rh} \leq  23.60 \;,\\
    \Delta N_\text{CMB} = 52.5 & \;\longrightarrow\; 8.39 \leq \Delta \tilde N_\text{rh} \leq  13.60\;,\\
    \Delta N_\text{CMB} = 54 & \;\longrightarrow\; 2.39 \leq \Delta \tilde N_\text{rh} \leq 7.60 \;,
\end{aligned}
\end{equation}
meaning that, e.g., for the case $\Delta N_\text{CMB}=50$ there is a model (i.e. a point in the parameter space of Fig.~\ref{fig:InstReh_Chi0D}) for which the choice $\Delta N_\text{CMB}=50$ corresponds to a reheating phase that lasted $\Delta \tilde N_\text{rh}  = 18.39$, one for which this choice corresponds to $\Delta \tilde N_\text{rh}  = 23.60$, and all the remaining models have $18.39< \Delta \tilde N_\text{rh} < 23.60$. 

The ranges in Eq.~\eqref{eq:reheating duration benchmark} show that we study different post-inflationary scenarios, ranging from very brief reheating, corresponding to $\Delta N_\text{CMB}=54$, to more extended phases, e.g. $\Delta N_\text{CMB}=50$. 
This will allow us to assess the impact of reheating, both on the large-scale constraints and the predictions on small scales. 

\subsection{CMB anisotropies constraints}
\label{sec: CMB constraints}

CMB measurements on large scales tightly constrain the amplitude and scale dependence\footnote{We defer the discussion of observational constraints on primordial non-Gaussianity to Sec.~\ref{sec:Non-Gaussianities}.} of the primordial scalar power spectrum. 
When comparing the predictions of an inflationary model with large-scale observations, it is useful to model $\mathcal{P}_\zeta(k)$ as
\begin{equation}
\label{eq:Pz power law}
    \mathcal{P}_\zeta(k) = \mathcal{A}_s \left( \frac{k}{k_\text{CMB}}\right)^{n_s-1 + \frac{\alpha_s}{2} \ln{(k/k_\text{CMB})}} \;, 
\end{equation}
where we omit higher-order contributions. 
The power spectrum tilt and its running are defined as 
\begin{equation}
    \label{eq:ns and alpha s definition}
    n_s-1\equiv \frac{\mathrm{d} \ln \mathcal{P}_\zeta }{\mathrm{d}\ln(k/k_\text{CMB})} \Big|_{k=k_\text{CMB}} \quad \text{and}\quad \alpha_s\equiv \frac{\mathrm{d}^2\ln \mathcal{P}_\zeta}{\mathrm{d}\ln(k/k_\text{CMB})^2} \Big|_{k=k_\text{CMB}} \;.
\end{equation}
Cosmological $\alpha$-attractors have \textit{universal} predictions for the large-scale observables, i.e. in the large $\Delta N_\text{CMB}$ limit they do not depend on the details of the potential shape~\cite{Kallosh:2013hoa, Kallosh:2013yoa, Achucarro:2017ing}.
Large perturbations can be produced within single- and multi-field $\alpha$-attractors, see e.g. Ref.~\cite{Iacconi:2021ltm} where this is shown for the case of exponential models. 
In presence of a peak in $\mathcal{P}_\zeta(k)$, the universal predictions get modified~\cite{Dalianis:2018frf,Iacconi:2021ltm}.
As a consequence of this modification, models which yield peaks at frequencies corresponding to the observational range of current and planned GWs observatories have $n_s$ too low to comply with current CMB bounds~\cite{Iacconi:2021ltm}.
When accommodated within hybrid inflation, the inflaton ($\phi$) potential is uplifted by the potential of the second field ($\chi$), and in the limit of large uplift the values of $n_s$ can be made larger, with $n_s\to1$~\cite{Kallosh:2022ggf}.
The presence of this second attractor for $n_s$ in the large uplift limit is the reason why the hybrid $\alpha$-attractor models we study here can be made consistent with CMB measurements of $n_s$, while also predicting peaks at phenomenologically interesting scales~\cite{Braglia:2022phb}. 
This is confirmed by our results obtained from scanning models in the $(\chi_0,\,d)$ parameter space, as we show below.   

Upon fixing $\Delta N_\text{CMB}=\{50, 52.5, 54\}$ (see the end of Sec.~\ref{sec: Delta N CMB}), for each model in the $(\chi_0,\,d)$ parameter space, we obtain $n_s$ and $\alpha_s$ by fitting the numerically obtained $\mathcal{P}_\zeta(k)$ at $0.04\leq k/\text{Mpc}^{-1}\leq 0.06$ with the expression~\eqref{eq:Pz power law}. 

\begin{figure}
\centering
\captionsetup[subfigure]{justification=centering}
\begin{subfigure}[b]{0.49\textwidth}
\includegraphics[width=\textwidth]{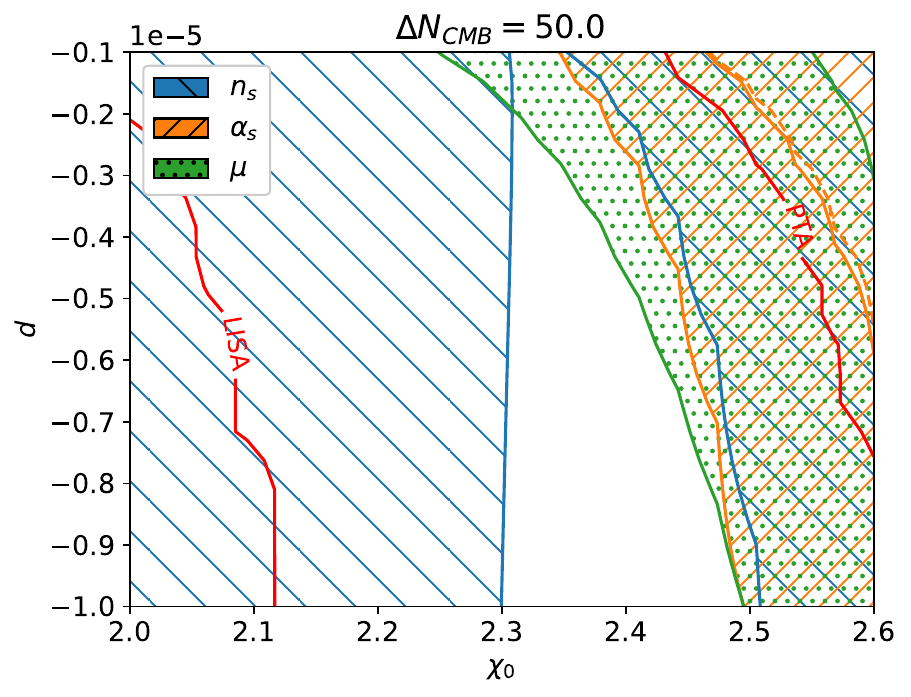}
\end{subfigure}
\begin{subfigure}[b]{0.49\textwidth}
\includegraphics[width=\textwidth]{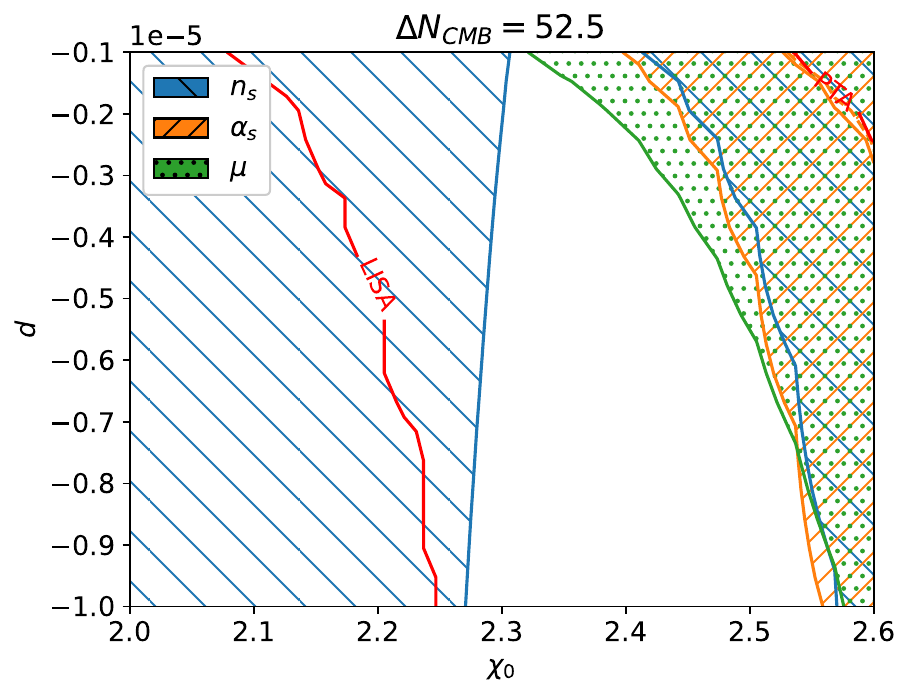}
\end{subfigure}
\begin{subfigure}[b]{0.49\textwidth}
\includegraphics[width=\textwidth]{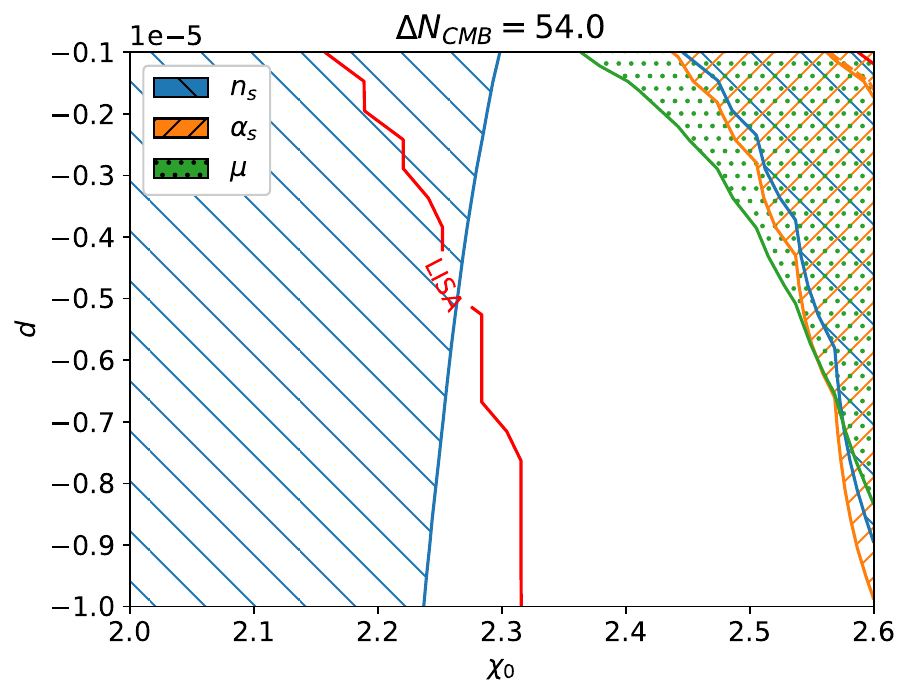}
\end{subfigure}
\caption{Large-scale constraints on the $(\chi_0,\, d)$ parameter space for different values of $\Delta N_\text{CMB}$, with fixed $\{\alpha=1,\, \tilde m = 0.3, \, \tilde g = 0.8\}$.
The white areas are those consistent with both CMB and $\mu$-constraints at least at $95\%$ C.L., see Sec.~\ref{sec: CMB constraints} and~\ref{sec: mu dist constraints} for more details. 
The red contours identify models whose scalar power spectrum peaks at the LISA and PTA pivot scale, $k_\text{LISA} = 3\times 10^{11}\, \text{Mpc}^{-1}$ and $k_\text{PTA}=10^7\, \text{Mpc}^{-1}$ respectively}. 
\label{fig:large scale constraints}
\end{figure}

Scalar perturbations produced during inflation are always accompanied by primordial gravitational waves, i.e. quantum vacuum fluctuations of tensor metric perturbations, stretched on super-horizon scales by the inflating background. 
Pending their detection, large-scale constraints on the amplitude of the primordial tensor power spectrum, $\mathcal{P}_\gamma$, are usually given relative to $\mathcal{P}_\zeta$, in terms of the tensor-to-scalar ratio $r\equiv\mathcal{P}_\gamma/\mathcal{P}_\zeta$.
For each model in the parameter space, we numerically\footnote{We note one cannot employ the single-field slow-roll prediction, $r=16\epsilon_1 + \mathcal{O}(\epsilon^2)$, since some of the models we consider display multi-field behavior already at CMB scales. We have explicitly checked the percent relative deviation of the first-order, slow-roll prediction and the numerical $r$. For fixed $\Delta N_\text{CMB}$, we find that the relative error increases towards the top-right corner of the $(\chi_0,\,d)$ parameter space. When varying $\Delta N_\text{CMB}$ for a fixed model, the percent relative deviation is inversely proportional to $\Delta N_\text{CMB}$. This reflects the fact that lowering $\Delta N_\text{CMB}$ pushes the horizon crossing of $k_\text{CMB}$ closer and closer to the peak scales, where multi-field effects are clearly relevant. Similar considerations also hold for the slow-roll expressions for $n_s$ and $\alpha_s$.} evaluate $r$ at $k=k_\text{CMB}$. 
We find that, for our choices of $\Delta N_\text{CMB}$, no model predicts $r$ above the $95\%$ C.L. observational upper bound derived from \textit{Planck} 2018 and BICEP/Keck 2018 data~\cite{Planck:2018nkj,Planck:2019nip,BICEP:2021xfz, Paoletti:2022anb}. 
The largest (smallest) value we find, $r=0.016$ ($r=2.255\times 10^{-7}$), corresponds to a model with $\Delta N_\text{CMB}=50$ ($\Delta N_\text{CMB}=50$) and $\{\chi_0=2.347,\,d=-10^{-6}\}$ ($\{\chi_0=2.6,\,d=-10^{-6}\}$). 
Since the observational upper bound on $r$ does not constrain the parameter space, there is no line in Fig.~\ref{fig:large scale constraints} corresponding to it.

We compare our results for $n_s$ and $\alpha_s$ with the measurements obtained for the $\Lambda\text{CDM}+r+\alpha_s$ cosmological model using \textit{Planck} 2018 and BICEP/Keck 2015 data~\cite{Planck:2018jri}, 
\begin{align}
\label{eq: LCDM+r+alpha_s constraints}
    n_s &=  0.9639 \pm 0.0044 \quad (68\%) \; \text{C.L.} \;,  \\
    \alpha_s &= -0.0069 \pm  0.0069 \quad (68\%) \; \text{C.L.} \;.  
\end{align}
In Fig.~\ref{fig:large scale constraints} we represent the parameter space constraints stemming from the measurements above.
In particular, the white regions are those complying with both constraints at least at $95\%$ C.L..
The red contours correspond to models whose $\mathcal{P}_\zeta(k)$ peaks exactly at the LISA and PTA pivot scales (see caption of Fig.~\ref{fig:large scale constraints}), which are representative of the frequency bands the two experiments operate at. These lines guide the eye and show at a glimpse the approximate location of the power spectrum peak across the parameter space.
They are not exactly vertical in the $(\chi_0,\,d)$ plane, indicating that the peak position, while being strongly correlated with $\chi_0$, also mildly depends on $d$. 
By observing how the white area changes with $\Delta N_\text{CMB}$, we conclude that, assuming $w=0$, an extended reheating stage (i.e. smaller $\Delta N_\text{CMB}$) shrinks the available region of parameter space.
This is because lowering $\Delta N_\text{CMB}$ moves the peak in $\mathcal{P}_\zeta(k)$ closer to $k_\text{CMB}$, see how the red contours change from one panel to the other. 

Models with $\chi_0\lesssim 2.3$ are excluded due to their prediction for $n_s$, which is too red to comply with~\eqref{eq: LCDM+r+alpha_s constraints}. 
For larger values of $\chi_0$, the $\alpha_s$ and $n_s$ constraints are comparable, with the bounds from $\alpha_s$ marginally stronger.
For these models, $n_s$ is either too close to unity or positive, predicting respectively a spectrum too close to scale invariance or blue. 
This is because the models in the top right portion of parameter space already display multi-field behavior at CMB scales.  

\subsection[\texorpdfstring{CMB ${\mu}$-distortion constraints}{CMB mu-distortion constraints}]{CMB $\bm{\mu}$-distortion constraints}
\label{sec: mu dist constraints}

Measurements of the CMB anisotropies constrain the primordial power spectrum at scales $0.0005 \lesssim k/\text{Mpc}^{-1}\lesssim 0.2$. 
A different type of CMB measurement, in particular of distortions of its energy spectrum from that of a perfect blackbody, allow to place constraints on $\mathcal{P}_\zeta(k)$ on scales smaller than those tightly constrained by CMB anisotropies~\cite{Zeldovich:1969ff, Hu:1994bz, Chluba:2011hw, Chluba:2012gq, Chluba:2015bqa, Schoneberg:2020nyg}.
There are different types of spectral distortions, among which the $\mu$-type constrain $\mathcal{P}_\zeta(k)$ up to $k\sim 10^4\, \text{Mpc}^{-1}$. 
They will be the focus of our analysis. 
See e.g. Refs.~\cite{Gow:2020bzo, Unal:2020mts} for an example of how upper bounds on $\mu$-type spectral distortions can be used to constrain the (possible) growth of $\mathcal{P}_\zeta(k)$. 

The total $\mu$-distortion induced by the scalar perturbations can be approximated by~\cite{Pajer:2012vz, Chluba:2015bqa}
 \begin{equation}
 \label{eq:mu equation}
     \mu \approx \int_{k_\text{min}}^{\infty} \frac{\mathrm{d}k}{k} \mathcal{P}_\zeta(k) W_{\mu}(k) \;,
 \end{equation}
with the $k$-space window function\footnote{We note that to fully model not only $\mu$- and $y$-type distortions, but also the complexity of non-$\mu$ and non-$y$ residual components, one could replace the analytic window function in Eq.~\eqref{eq: mu window} with a numerical counterpart, see e.g. Refs.~\cite{Cyr:2023pgw, Tagliazucchi:2023dai} for more details. 
While the two can yield different $\mu$ values when the power spectrum peaks at the edges of $W_{\mu}(k)$, for many different spectral shapes of $\mathcal{P}_\zeta(k)$ they are shown to yield consistent results (see e.g. Fig.~4 in Ref.~\cite{Tagliazucchi:2023dai}). 
For this reason, we employ Eq.~\eqref{eq: mu window} in this work, and leave for the future the exploration of effects from numerical $W_\mu(k)$.}
 \begin{equation}
 \label{eq: mu window}
     W_{\mu}(k) \approx 2.27 \left\{\exp \left[-\left(\frac{\hat{k}}{1360}\right)^2 \Bigg/\left(1+\left(\frac{\hat{k}}{260}\right)^{0.3} + \frac{\hat{k}}{340}\right)\right] - \exp \left[-\left(\frac{\hat{k}}{32}\right)^2\right]\right\},
 \end{equation}
where $\hat{k}=k/1\,\text{Mpc}^{-1}$ and $k_\text{min} \simeq 1 \text{Mpc}^{-1}$. 

By using the numerical $\mathcal{P}_\zeta(k)$ obtained with \texttt{PyTransport}, we compute the integral in Eq.~\eqref{eq:mu equation}. 
In particular, we choose an effective upper limit in the integral which ensures convergence of the $\mu$ value. 
We then compare these results with the current observational upper bound, $\mu < 9\times 10^{-5}$ ($95\%$ C.L.), obtained from COBE/FIRAS observations~\cite{Fixsen:1996nj}, and display in Fig.~\ref{fig:large scale constraints} the area of parameter space which is excluded by it.
For fixed $\Delta N_\text{CMB}$, models are excluded in the top-right area of the parameter space, as the power spectrum growth on large scales which these models display is not compatible with COBE/FIRAS observations. 
As for the case of $n_s$ and $\alpha_s$ constraints, the duration of reheating impacts greatly the viability of models. 
This can be seen by comparing how the portion of parameter space excluded by $\mu$ constraints changes with $\Delta N_\text{CMB}$. 
Smaller $\Delta N_\text{CMB}$ values move the large scales towards the peak in $\mathcal{P}_\zeta(k)$, and larger portions of the parameter space are ruled out. 
We note that for $\Delta N_\text{CMB}=50$ the models in the top-right corner of the parameter space, while displaying multi-field behavior on large scales, are not sufficiently enhanced to violate the $\mu$ constraint. 

By combining $n_s$ and $\alpha_s$ measurements together with $\mu$-distortions constraints, we obtain the area of parameter space which is consistent with large-scale tests. 
For viable models, the peak in $\mathcal{P}_\zeta(k)$ is located between PTA and LISA scales ($\Delta N_\text{CMB}=50$), or closer to LISA scales ($\Delta N_\text{CMB}=\{52.5, \, 54\}$), see the red lines in Fig.~\ref{fig:large scale constraints}.
For this reason, in Sec.~\ref{sec:GW} we consider the capability of LISA to test the scalar-induced GWs.

Before moving on, let us underline that the peak positions and the results displayed in Fig.~\ref{fig:large scale constraints} depend on the reduced parameter space we work with. 
For example, while a peak at PTA scales is not compatible with large-scale constraints in Fig.~\ref{fig:large scale constraints}, this conclusion might change when including the effect of the parameters which have been fixed here. 

\section{Non-Gaussianity and perturbativity}
\label{sec:Non-Gaussianities}
In Sec.~\ref{sec:large-scale tests} we have focused on the $2$-point function of the comoving curvature perturbation and its constraints on large scales. While this description of the statistics of $\zeta$ is exhaustive for SFSR inflation~\cite{Maldacena:2002vr}, when more fields play a role small deviations from a Gaussian distribution could appear.
The first diagnostic tool is the 3-point correlation function, whose Fourier transform is  
\begin{equation}      
\label{eq:def 3 pt corr funct}\left\langle\zeta(\mathbf{k_1})\zeta(\mathbf{k_2})\zeta(\mathbf{k_3})\right\rangle=(2\pi)^3 \delta^3(\mathbf{k_1}+\mathbf{k_2}+\mathbf{k_3})B_{\zeta}(k_1,k_2,k_3)\;, 
\end{equation} 
where $B_\zeta$ is referred to as the \textit{bispectrum}.
The amplitude and shape-dependence of primordial non-Gaussianities (PNGs) depends on the physical mechanism that produced them, and can hence be used to classify and test inflationary models (for a review on the types of PNGs and their origin, see, e.g., Refs.~\cite{Bartolo:2004if,Liguori:2010hx,Chen:2010xka,Komatsu:2010hc,Renaux-Petel:2015bja}). 
The latest large-scale constraints on equilateral, local and orthogonal PNGs are placed by the \textit{Planck} survey~\cite{Planck:2019kim}, and allow us to rule out models that overproduce them at CMB scales.
Observational constraints are usually placed on the reduced bispectrum, defined as  
\begin{equation}
\label{eq:fNL}
    f_\text{NL}(k_1,k_2,k_3)=\frac{5}{6}\frac{B_{\zeta}(k_1,k_2,k_3)}{P_{\zeta}(k_1)P_{\zeta}(k_2)+P_{\zeta}(k_1)P_{\zeta}(k_3)+P_{\zeta}(k_2)P_{\zeta}(k_3)} \;, 
\end{equation}
where $P_\zeta(k)=(2\pi^2 /k^3) \mathcal{P}_\zeta(k)$. 

We start by computing the amplitude of $f_\text{NL}(k_1,k_2,k_3)$ for four different triangle configurations, obtained for a model with $\Delta N_\text{CMB}=50$ and parameters
\begin{equation}
\label{eq:baseline_fNL}
    \alpha = 1, \quad \Tilde{g} = 0.8, \quad \Tilde{m} = 0.3, \quad \chi_0 = 2.375, \quad d = -6.6173\times10^{-6} \;. 
\end{equation}
This specific model agrees with the large-scale constraints discussed in Sec.~\ref{sec:large-scale tests}, and presents a large peak in $\mathcal{P}_\zeta(k)$ without overproducing PBHs (see Sec.~\ref{sec: PBH}).
We numerically compute the value of the reduced bispectrum by using \texttt{PyTransport}, and parameterize the three wavenumbers $k_1$, $k_2$ and $k_3$ in Eq.~\eqref{eq:fNL} in terms of $\alpha$, $\beta$ and the total scale $k_t=k_1 + k_2 + k_3$, according to 
\begin{equation}
\label{eq:alpha and beta def}
k_1=k_t\frac{(1-\beta)}{2}, \quad k_2=k_t\frac{(1+ \alpha +\beta)}{4}, \quad k_3=k_t\frac{(1-\alpha+\beta)}{4} \;. 
\end{equation}
We consider four different configurations, namely
\begin{itemize}
    \item equilateral, with $k_1=k_2=k_3=k_t/3$ (or equivalently $\alpha=0$ and $\beta=1/3$);
    \item folded, with $k_1=k_t/2,\,  k_2=k_3=k_t/4$ (or $\alpha = \beta = 0$);
    \item an isosceles, with $k_1=k_t/6,\, k_2=k_3=5k_t/12$ (or $\alpha = 0$ and $\beta = 2/3$);
    \item a squeezed, with $k_1=k_t/20, \,k_2=k_t/2, \, k_3=9k_t/20$ (or $\alpha = 0.1$ and $\beta = 0.9$). Note that all three modes $k_1$, $k_2$ and $k_3$ are comparable, with the maximum squeezing given by $k_1/k_2=0.1$.
\end{itemize}   
\begin{figure}
\begin{center}
{\includegraphics[width=.7\linewidth]{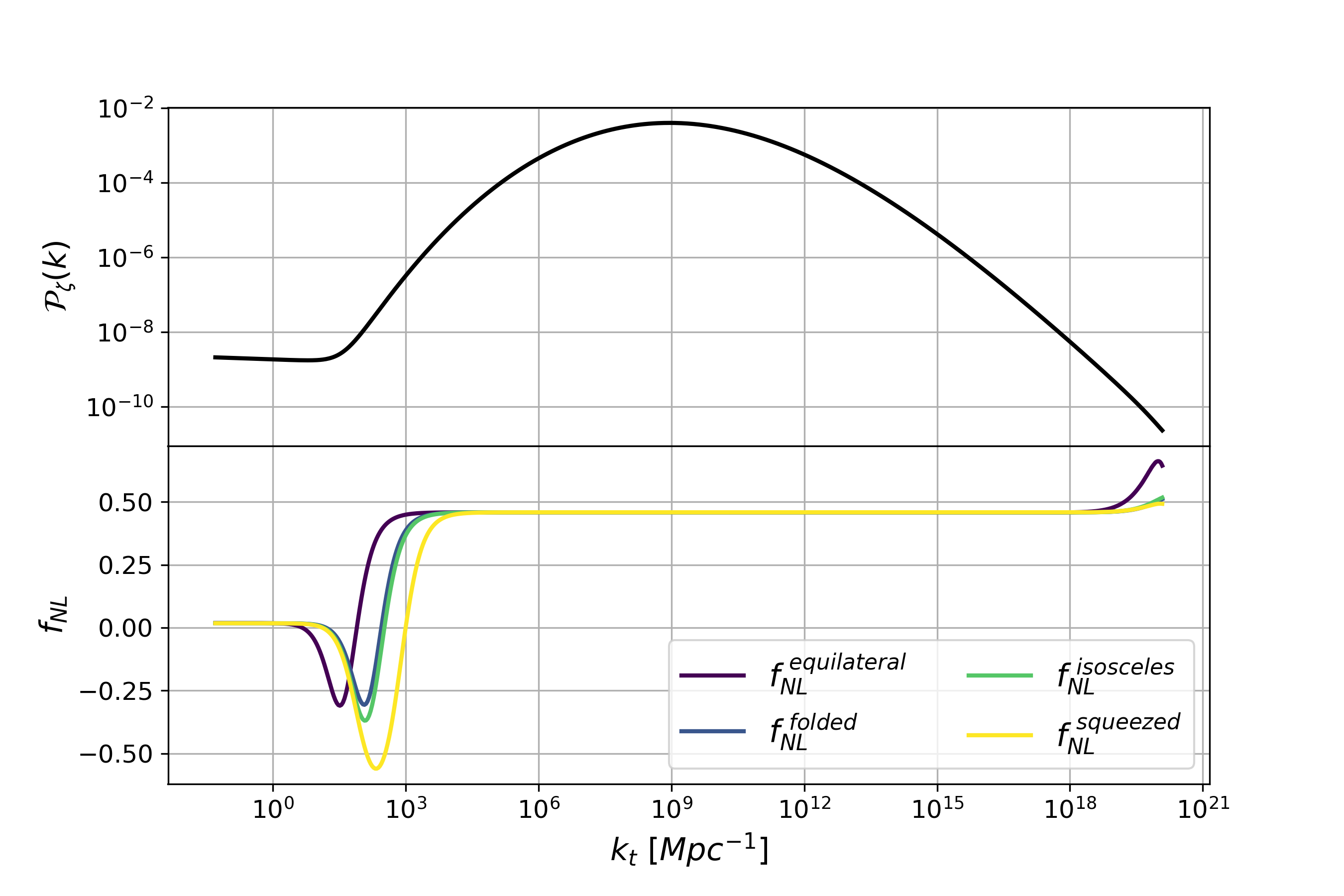}}
\caption{Scale-dependence of the dimensionless scalar power spectrum (\textit{top}) and reduced bispectrum $f_\text{NL}$ (\textit{bottom}) for a model with parameters~\eqref{eq:baseline_fNL}. 
See the main text for more details on the triangle configurations considered.}
\label{fig:fNLConf}
\end{center}
\end{figure}
In Fig.~\ref{fig:fNLConf} the results are represented as a function of $k_t$, with $k_t$ ranging from the CMB pivot scale to the one that crossed the horizon at the end of inflation.
First, we note that on large scales the reduced bispectrum is small, e.g. ${f_\text{NL,\,sq}}(k_t=k_\text{CMB})=0.01768$, well within $\textit{Planck}$ constraints~\cite{Planck:2019kim}, regardless of the configuration. 
At the pivot scale, the variation in the value of $f_\text{NL}$ between the four configurations considered is less than $4\%$. 
The value of $\fnl$ in the squeezed configuration when $k_t$ is a long mode is in agreement with the prediction from Maldacena's consistency relation~\cite{Maldacena:2002vr}, $f_\text{NL, CR} (k_t=k_\text{CMB})=(5/12)(1-n_\text{s})=0.01762$.
This is due to the fact that, when the large scales crossed the horizon, the dynamics of the model is essentially SFSR, driven by the slowly-rolling field $\phi$, and the long mode $k_1\ll k_2,k_3$ is adiabatic~\cite{Mooij:2015yka, Bravo:2017wyw}.
While the results represented in Fig.~\ref{fig:fNLConf} are for a single model, we have computed the (equilateral and squeezed) reduced bispectrum on large scales over the $(\chi_0, \, d)$ parameter space compatible with the large-scale constraints discussed in Sec.~\ref{sec:large-scale tests}. 
We found $f_\text{NL,\,eq}(k_t=3k_\text{CMB})=\left[0.012, 0.019\right] $ and $f_\text{NL,\,sq}(k_t=3k_\text{CMB})=\left[0.009, 0.019\right] $, compatible with CMB constraints~\cite{Planck:2019kim}. 
Hence, the computation of the bispectrum on large scales does not allow us to further constrain the parameter space. 
It is important to point out that this study of non-Gaussianity for models compatible with large-scale constraints does not depend on the duration of reheating, i.e. the specific chosen value of $\Delta N_\text{CMB}$. 
All the models in the $(\chi_0, d)$ parameter space region compatible with the large-scale constraints on $n_\text{s}$, $\alpha_s$ and $\mu$-distortion (see Fig.~\ref{fig:large scale constraints}), will have negligible non-Gaussianity on CMB scales regardless of $\Delta N_\text{CMB}$. 
This is because models complying with large-scale constraints effectively behave as in SFSR inflation at those scales.

Moving towards smaller scales, one observes in Fig.~\ref{fig:fNLConf} that the reduced bispectrum is constant for modes that correspond to the peak in the scalar power spectrum. 
This applies regardless of the triangle configuration, and the $\fnl$ plateau is configuration-independent. 
The small amplitude of $\fnl$, e.g. $f_\text{NL}(k_t=k_\text{peak})=0.45883$, suggests that non-linearities due to cubic interactions are subdominant and do not pose a threat to perturbativity for this model, see Eq.~\eqref{eq:ratio 1loop over tree} and the related discussion. 
Note that Maldacena's consistency relation would not reproduce the numerical squeezed bispectrum when $k_t$ is a short mode as the long mode of the configuration, $k_1\ll k_2,k_3$, is not adiabatic.

The scale- and shape-independence of $f_\text{NL}$ over the peak scales observed in Fig.~\ref{fig:fNLConf} suggests that the bispectrum is of the local type.
\begin{figure}
\begin{center}
{\includegraphics[width=.7\linewidth]{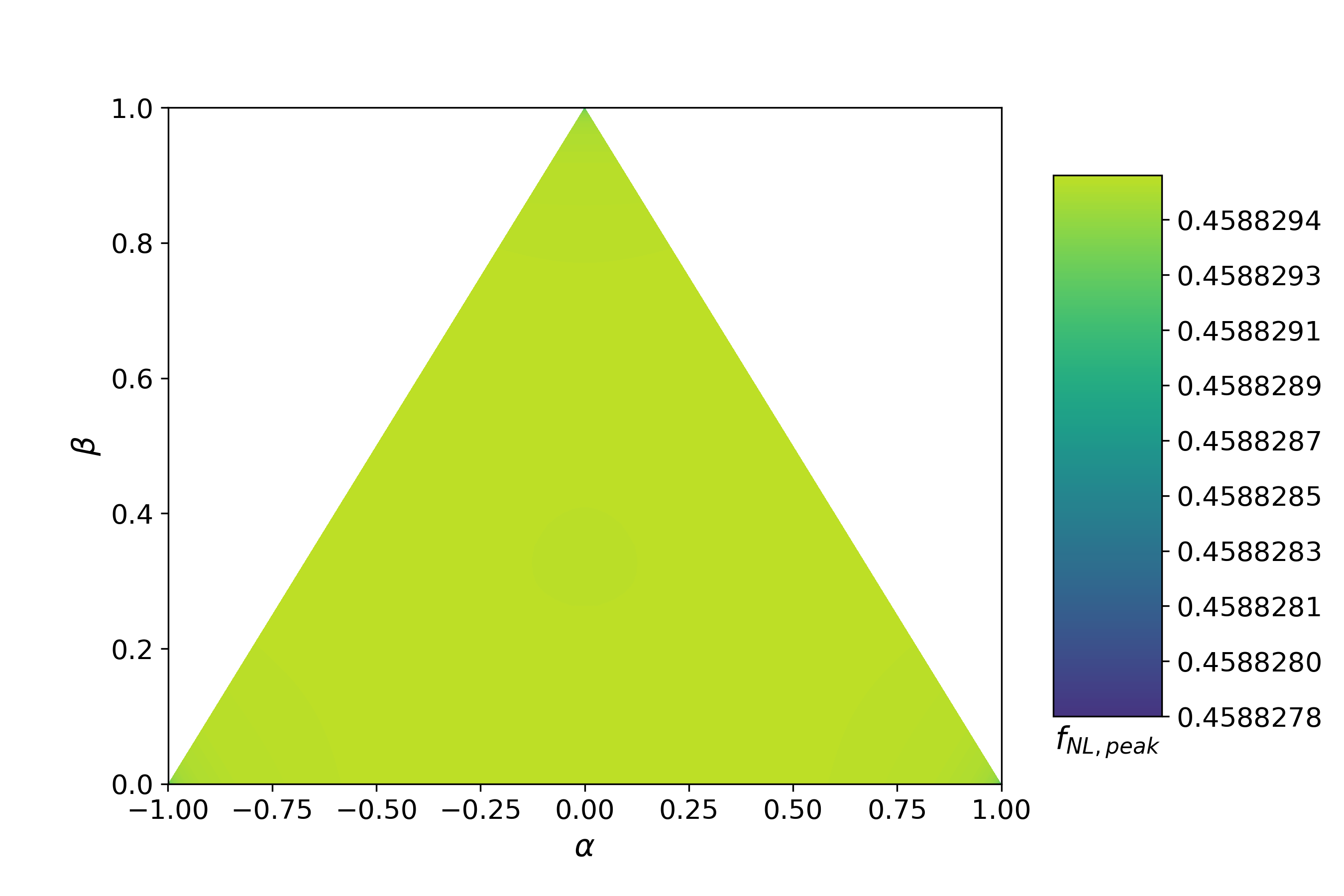}}
\caption{Shape-dependence of the reduced bispectrum $f_\text{NL}(k_1,\, k_2,\, k_3)$ at peak scales for the model with parameters~\eqref{eq:baseline_fNL}. 
See Eq.~\eqref{eq:alpha and beta def} for the definition of $\alpha$ and $\beta$. 
Note that the squeezed configuration corresponds to the points $(\alpha, \beta)=\{(-1,\, 0),\, (1,\, 0),\, (0,\, 1)\}$ , while the equilateral configuration corresponds to $(\alpha, \beta) = (0,
\, 1/3)$.}
\label{fig:fNLTriangle}
\end{center}
\end{figure}
To test this over a much larger number of configurations, we compute $f_\text{NL}$ for $k_t = 3 k_\text{peak}$ as a function of $\alpha$ and $\beta$, see Eq.~\eqref{eq:alpha and beta def}, and represent the results in Fig.~\ref{fig:fNLTriangle}.
This shows that the reduced bispectrum is almost constant for all configurations, confirming that over the peak scales it is of the local type. 
We note that we have sampled the $(\alpha,\, \beta)$ space such that, in the vicinity of points corresponding to squeezed triangles, all three modes are selected within the $\fnl$ plateau. 

After establishing for the model with parameters~\eqref{eq:baseline_fNL} that the bispectrum at peak scales is of the local type and has small amplitude, $\fnl \sim \mathcal{O}(0.1)$, we move on to extend this study to all models in the $(\chi_0,\,d)$ parameter space compatible with large-scale constraints. 
\begin{figure}
\begin{center}
{\includegraphics[width=.7\linewidth]{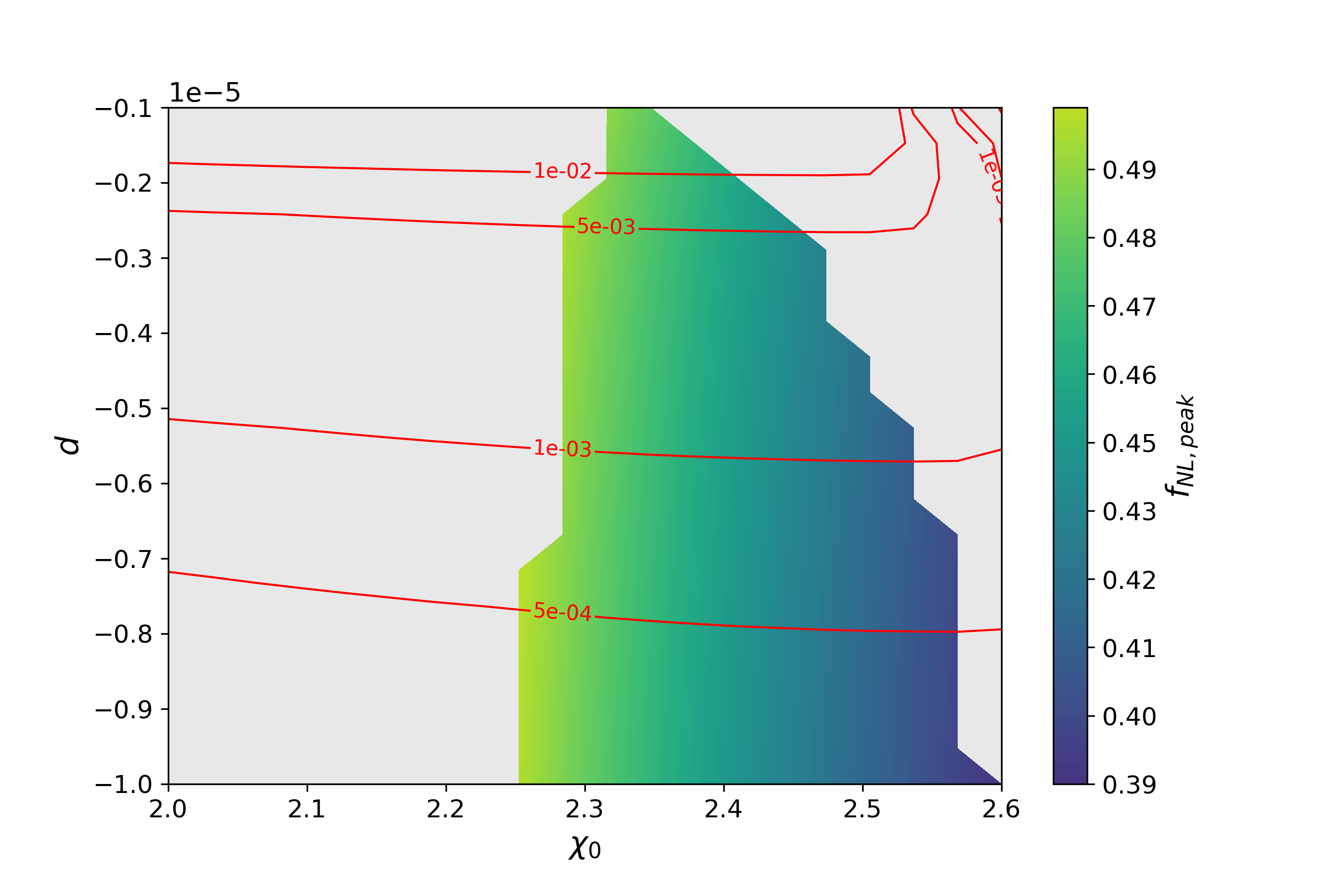}}
\caption{Value of the reduced bispectrum in the equilateral configuration at peak scales, $f_\text{NL}(k_\text{peak})$. 
This is computed over the region of the $(\chi_0,d)$ parameter space in agreement with the large-scale constraints discussed in Sec.~\ref{sec:large-scale tests} for  $\Delta N_\text{CMB}=54$. 
The red lines represent contours of constant $\mathcal{P}_{\zeta}(k_\text{peak})f_\text{NL}(k_\text{peak})^2$, which we use to estimate the size of the one-loop correction to the power spectrum due to cubic interactions.}
\label{fig:fNLChi0D}
\end{center}
\end{figure}
The results for models with $\Delta N_\text{CMB}=54$ are represented in Fig.~\ref{fig:fNLChi0D}. 
Here, we first note that $\fnl$ mainly depends on the value of $\chi_0$, i.e. is more correlated with the position of the peak in the scalar power spectrum rather than with its amplitude.  
Nevertheless, this dependence is mild and $\fnl$ always remains of the same order of magnitude, $f_\text{NL}(k_\text{peak})=[0.39 ,\, 0.50]$. 
We find that $f_\text{NL} \propto \chi_0^{-1.8}$, close to the scaling found in Refs.~~\cite{Abolhasani:2010kn,Mulryne:2011ni}, $f_\text{NL} \propto \chi_0^{-2}$. 

We now apply the results found for the bispectrum to the issue of perturbativity, see e.g. the analysis in Ref.~\cite{Iacconi:2023slv}.
For a model with local non-Gaussianity, the curvature perturbation can be approximated as (see Ref.~\cite{Wands:2010af})\footnote{The expansion in  Eq.~\eqref{eq:zetaLoop} is valid only if the curvature perturbation $\zeta(\mathbf{x})$ is a function of a single Gaussian field, $\zeta_\text{G}(\mathbf{x})$ in this case. While we note that this could in principle not be the case for multi-field models of inflation, it is for the hybrid models under analysis. This is reflected in the numerical $\fnl$ results, which are consistent with local non-Gaussianity. For completeness, we have also performed a numerical $\delta N$ computation akin to the one in Ref.~\cite{Iacconi:2023slv}, showing that the peak in the power spectrum is mainly sourced by fluctuations of the $\chi$ field.}
\begin{equation}
\label{eq:zetaLoop}
    \zeta(\textbf{x}) = \zeta_\text{G}(\textbf{x}) + \frac{3}{5} f_\text{NL} \left(\zeta_\text{G}(\textbf{x})^2-\langle\zeta_\text{G}(\textbf{x})^2\rangle \right) \;, 
\end{equation}
where $\zeta_\text{G}(\mathbf{x})$ is a Gaussian field and $\left\langle\zeta_\text{G}(\textbf{x})^2\right\rangle$ its variance.
For scale-invariant $\mathcal{P}_{\zeta_\text{G}}$~\cite{Lyth:2006gd,Kumar:2009ge}, the ratio between the 1-loop correction due to the non-linear component of Eq.~\eqref{eq:zetaLoop} and the tree-level, Gaussian power spectrum is 
\begin{equation}
    \label{eq:ratio 1loop over tree}
    \frac{\mathcal{P}_\zeta^{\text{1-loop}}}{\mathcal{P}_{\zeta_\text{G}}} \approx {f_\text{NL}}^2 \mathcal{P}_{\zeta_\text{G}} \;. 
\end{equation}
This goes to show that the value of $f_\text{NL}$ in itself is not a good indicator of the validity of the perturbative approach. 
Although the hybrid models under analysis do not present a scale-invariant power spectrum (rather they display a peak), the magnitude of Eq.~\eqref{eq:ratio 1loop over tree} can be still used as an approximate indicator of whether non-linearities --in particular those due to cubic interactions-- are important.  
In Fig.~\ref{fig:fNLChi0D} we display contours of constant $f_\text{NL}(k_\text{peak})^2 \mathcal{P}_{\zeta_\text{G}}(k_\text{peak})$, where $\mathcal{P}_{\zeta_\text{G}}$ is the tree-level power spectrum, i.e. the one obtained by solving the linear equation of motion for $\zeta$. 
These are always well below unity, and 
therefore indicate that non-linear corrections to the tree-level power spectrum at peak scales are subdominant. 
This is an important point, in fact one of the main results of our analysis of hybrid $\alpha$-attractors. 
Fig.~\ref{fig:fNLChi0D} shows that hybrid $\alpha$-attractors behave differently from multi-field polynomial $\alpha$-attractors~\cite{Braglia:2020eai, Kallosh:2022vha}. 
Indeed, for the latter the peak is produced by geometrical effects and perturbativity is violated at peak scales~\cite{Iacconi:2023slv}. 

Pending a more systematic check of perturbativity, for now the smallness of non-linear corrections to the scalar power spectrum justify the use of the tree-level $\mathcal{P}_\zeta$ to study the phenomenology of hybrid $\alpha$-attractors on small scales.

\section{Small-scale tests}
\label{sec: small scale test}
In this Section we implement current and future small-scale tests for the hybrid $\alpha$-attractor model~\eqref{eq:hybrid potential}.
In Sec.~\ref{sec: PBH} we compute the total amount of PBHs produced as a fraction of cold dark matter, and use this to further constrain the parameter space.  
Sec.~\ref{sec:GW} is devoted to the study of scalar-induced GWs. 
Motivated by the typical scales at which $\mathcal{P}_\zeta$ peaks for models in the $(\chi_0,\,d)$ parameter space which are compatible with large-scale constraints (see Fig.~\ref{fig:large scale constraints}), we focus on the capability of LISA and ET to test these models in the near future.

\subsection{Primordial black hole formation}
\label{sec: PBH}
A large peak in the primordial scalar power spectrum enhances the likeliness of large over-densities at horizon re-entry, and therefore can lead to primordial black hole (PBH) formation~\cite{Carr:1974nx}. 
Since the probability of forming PBHs depends strongly on how likely large perturbations are, the precise calculation of the PBHs abundance is highly sensitive to the amplitude of the power spectrum, as well as to any non-Gaussian tail of the probability distribution function (PDF). 
Non-perturbative non-Gaussianities are expected due to stochastic effects during inflation (see, e.g., the recent Review~\cite{Vennin:2024yzl} and references therein), and any perturbative expansion around a Gaussian distribution, such as the one in Eq.~\eqref{eq:zetaLoop}, would fail to capture the tail behavior of the PDF of $\zeta$. 
Nonetheless, modelling of stochastic inflation constitutes a whole research program by itself, and is therefore beyond the scope of this work.
Here, we include non-Gaussianities perturbatively. 
Motivated by the findings described in Sec.~\ref{sec:Non-Gaussianities}, in particular that $f_\text{NL}\sim \mathcal{O}(0.1)$, we assume that $\zeta$ follows a Gaussian distribution.  

In the following, we assume PBHs are formed from the collapse of over-dense regions during radiation domination, and apply the Press-Schechter methodology~\cite{Press:1973iz} to compute their abundance today\footnote{There are other methods to compute the PBH abundance, see e.g. Ref.~\cite{Young:2024jsu} for a recent Review.}.
This is defined as the fraction of cold dark matter in the form of PBHs today, 
\begin{equation}
\label{fPBH def}
    f_\text{PBH} \equiv \frac{\Omega_\text{PBH}}{\Omega_\text{CDM}} \;, 
\end{equation}
where $\Omega_i\equiv \rho_i/\rho_\text{tot}$ is the density parameter of the species $i$. 

Rather than working directly at the level of $\zeta(t,\mathbf{x})$, or the density contrast $\delta(t,\mathbf{x}) \equiv \left(\rho(t,\mathbf{x}) -\bar \rho\right) /\bar \rho$, when establishing whether a perturbation will collapse to form a PBH, we work with the compaction function, $\delta_{R}$, which can be thought of as a rescaled version of the density contrast smoothed with a top-hat window function over a sphere of radius $R$~\cite{Musco:2018rwt}. 
Technically, we will use a compaction\textit{-like} function, as we employ instead the modified Gaussian window function, whose Fourier transform is $\tilde W(k,R) = \exp{-(kR)^2/4}$. 
PBHs will form from regions
where the compaction $\delta_R$ is above some threshold value, $\delta_{R}^c$. Whilst the precise value of the threshold depends on the shape of the perturbation~\cite{Germani:2018jgr, Musco:2018rwt, Young:2019osy, Musco:2020jjb}, in this work we consider a unique threshold value (as done, e.g., in Ref.~\cite{Gow:2020bzo}), which we discuss below.

The primordial curvature perturbation, $\zeta$, and the compaction, $\delta_R$, are non-linearly related~\cite{Young:2019yug}.
Assuming spherical symmetry\footnote{We consider type I perturbations (see Ref.~\cite{Kopp:2010sh} for more details).}, on super-horizon the compaction $\delta_{R}$ is related to its linear component, $\delta_{R,l}$, itself being linearly related to $\zeta$, as
\begin{equation}
\label{non-linear compaction}
    \delta_R = \delta_{R,l} - \frac{3}{8}{\delta_{R,l}}^2 \;,
\end{equation}
where we assume radiation domination. 
The linear ${\delta_{R,l}}$ follows a Gaussian distribution, $P(\delta_{R,l})$, and its variance is computed from the primordial scalar power spectrum according to 
\begin{equation}
    \label{sigma_R^2}
    \sigma^2(R) = \int_0^\infty  \mathrm{d}\ln k \; \frac{16}{81}\, (kR)^4 \, \tilde W(kR)^2 \, \mathcal{P}_\zeta(k) \;.
\end{equation}
The PBH abundance is then given as
\begin{equation}
\label{fPBH in practice}
    f_\text{PBH} = \frac{2}{\Omega_\text{CDM}} \int \mathrm{d}\ln R \;\frac{R_\text{eq}}{R} \, \int^\infty_{\delta_{R,l}^c} \mathrm{d}\delta_{R,l} \; \frac{M_\text{PBH}}{M_H} P(\delta_{R,l})  \;,
\end{equation} 
where $R_\text{eq}$ is the horizon size at matter-radiation equality and $\delta_{R,l}^c$ is the critical value for the linear compaction, determined using Eq.~\eqref{non-linear compaction} and the value of $\delta_R^c$. 
We include the effect of critical collapse~\cite{Niemeyer:1997mt} by having  
\begin{equation}
    \label{critical collapse}
    \frac{M_\text{PBH}}{M_H} = \mathcal{K}  \left(\delta_R  - \delta_R^c \right)^\gamma \;,
\end{equation}
where we substitute Eq.~\eqref{non-linear compaction} and take $\mathcal{K}=10$, $\gamma=0.36$ and $\delta_R^c=0.25$, all values being coherent with our choice of window function~\cite{Young:2019osy, Young:2020xmk}. 
  
Following the procedure outlined above, we compute $f_\text{PBH}$ over the region of parameter space that is consistent with the large-scale constraints presented in Sec.~\ref{sec:large-scale tests}. 
This allows us to further constrain the parameter space by excluding regions which yield to PBH overproduction $(f_\text{PBH}>1)$, and to highlight areas that can lead to significant PBH production, which we define by the criterion $10^{-3}\leq f_\text{PBH}\leq 1$. 
\begin{figure}
\centering
\captionsetup[subfigure]{justification=centering}
\begin{subfigure}[b]{0.49\textwidth}
\includegraphics[width=\textwidth]{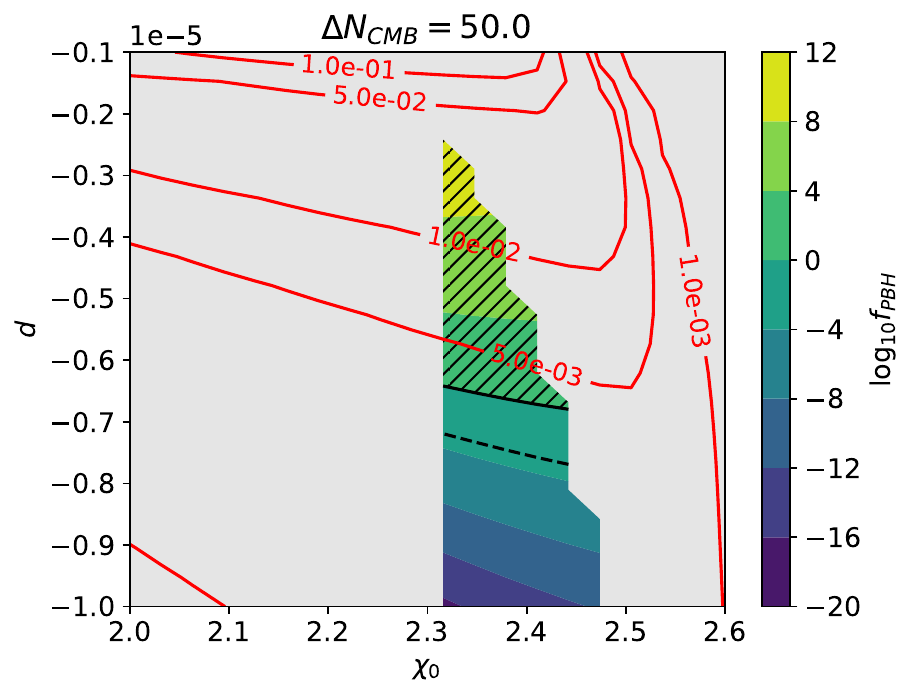}
\end{subfigure}
\begin{subfigure}[b]{0.49\textwidth}
\includegraphics[width=\textwidth]{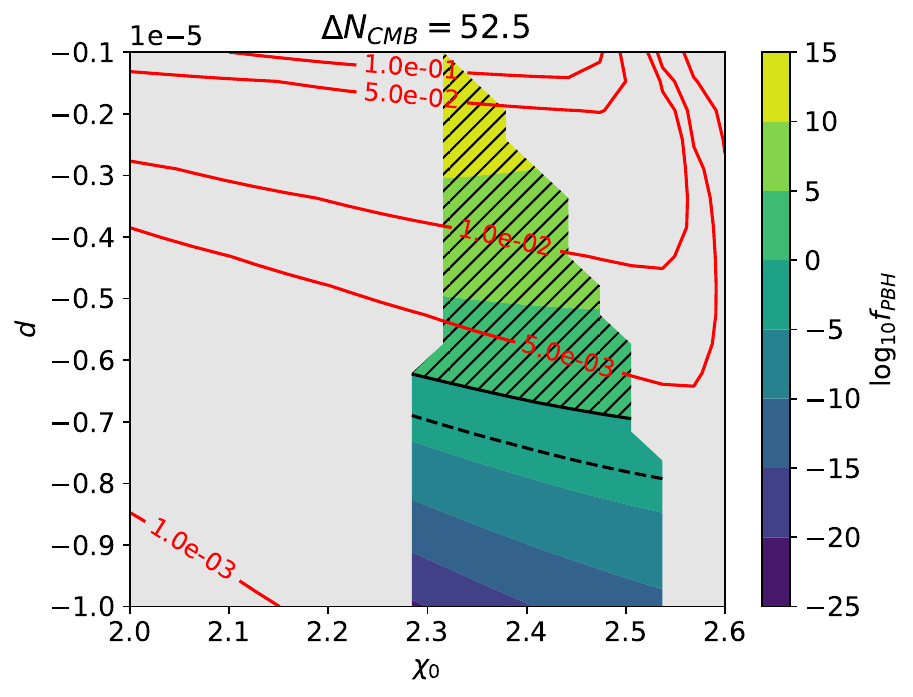}
\end{subfigure}
\begin{subfigure}[b]{0.49\textwidth}
\includegraphics[width=\textwidth]{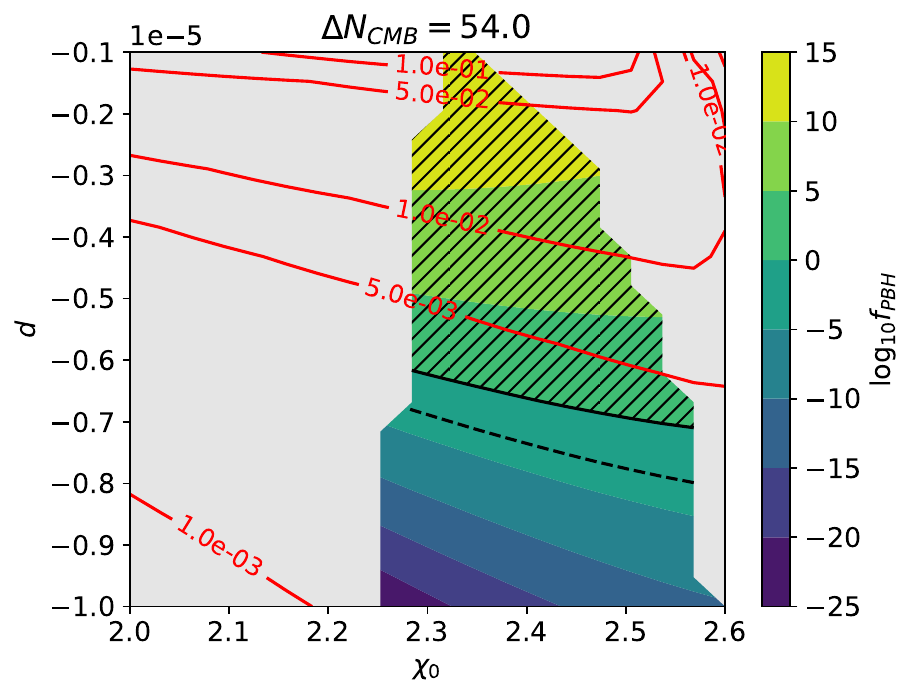}
\end{subfigure}
\caption{Value of $f_\text{PBH}$ computed over regions of the $(\chi_0,\,d)$ parameter space compatible with the large-scale constraints discussed in Sec.~\ref{sec:large-scale tests}. 
Each panel corresponds to a different value of $\Delta N_\text{CMB}$, see Sec.~\ref{sec: Delta N CMB} for more details on these choices. 
The area hatched with black dashes yields $f_\text{PBH}>1$, and therefore highlights models excluded due to PBH overproduction. 
The continuous and dashed black lines correspond to models yielding $f_\text{PBH}=1$ and $f_\text{PBH}=10^{-3}$ respectively. 
The superimposed red lines correspond to contours of constant $\mathcal{P}_\zeta(k_\text{peak})$.
}
\label{fig:fPBH}
\end{figure}
We represent our results in Fig.~\ref{fig:fPBH}, where each panel corresponds to a different choice for $\Delta N_\text{CMB}$. 
Our results show that imposing $f_\text{PBH}\leq1$ yields significantly stronger constraints on the parameter space with respect to the approximate criterion which is often used, $\mathcal{P}_\zeta(k_\text{peak})\lesssim 0.01$. 
Indeed, the precise value of $f_\text{PBH}$ is model dependent~\cite{Gow:2020bzo}, and a careful numerical computation is required in order to meaningfully constrain inflation.

While here we have imposed the theoretical requirement  $f_\text{PBH}\leq1$, one could further constrain the parameter space by means of experimental limits on the fraction of PBHs as a function of their mass, $M$. 
For a review see, e.g., Refs.~\cite{Sasaki:2018dmp, Carr:2020gox, Green:2024bam}.
In the reduced parameter-space slice we focus on, viable models display peaks between LISA and PTA scales when reheating is extended ($\Delta N_\text{CMB}=50$), and closer to LISA scales for short reheating stages ($\Delta N_\text{CMB}=\{52.5,\, 54\}$), see Fig.~\ref{fig:large scale constraints}. 
This roughly corresponds to PBHs produced with masses $10^{-15}\lesssim M/M_\odot \lesssim 1$ (see e.g. Fig.~6 of Ref.~\cite{Gow:2020bzo} or Fig.~3 of Ref.~\cite{Green:2024bam}), whose lifetime is longer than the age of the Universe. 
For this range of masses, experimental bounds on $f_\text{PBH}(M)$ affect models with significant PBH production, e.g. $10^{-2}\leq f_\text{PBH}\leq 1$ (see Fig.~4 of Ref.~\cite{Green:2024bam}).  
By inspecting Fig.~\ref{fig:fPBH}, one sees that applying these bounds in the present analysis will therefore only marginally strengthen the bounds obtained from $f_\text{PBH}\leq 1$.
On the other hand, including the effects of a multi-dimensional parameter space might produce models that, while being consistent with large-scale constraints, display peaks at different scales. 
For example, for $k_\text{peak}$ around ground-based interferometer scales experimental bounds on $f_\text{PBH}(M)$ would be more constraining than requiring $f_\text{PBH}\leq 1$ only. 
This is due to the fact that the PBHs produced in this case would be evaporating at a significant rate today or have evaporated completely, and their abundance is constrained up to very small values of $f_\text{PBH}(M)$ (with the specific values depending on $M$)~\cite{Carr:2020gox}.
Note that the capability for this type of analysis to constrain the parameter space will increase in the future, see Ref.~\cite{Bird:2022wvk}. 
Including experimental constraints on $f_\text{PBH}(M)$ is one of the many avenues along which our work can be extended in the future. 

 \subsection{Scalar-induced gravitational waves}
\label{sec:GW}

Tensor perturbations are sourced by scalar modes at second order in perturbation theory~\cite{Ananda:2006af, Baumann:2007zm, Domenech:2021ztg}.
We will refer to these as scalar-induced gravitational waves (SIGW).
Due to the smallness of scalar fluctuations produced within SFSR inflation, the SIGW signal is predicted to be too small for detection, see e.g. Ref.~\cite{Caprini:2015tfa}. 
However, if $\mathcal{P}_\zeta$ grows significantly larger, as in the models we study here, SIGW might be probed by future GW observatories, such as the space-based Laser Interferometer Space Antenna (LISA)~\cite{LISA:2017pwj, LISACosmologyWorkingGroup:2022jok}, Taiji~\cite{Hu:2017mde} and TianQin~\cite{TianQin:2015yph}, and the third-generation ground-based detectors Einstein Telescope (ET)~\cite{Punturo:2010zz, Maggiore:2019uih} and Cosmic Explorer~\cite{Reitze:2019iox}.
On larger scales, analysis of the most recent PTA data by the NANOGrav~\cite{NANOGrav:2023gor, NANOGrav:2023hde}, EPTA and InPTA~\cite{EPTA:2023fyk, EPTA:2023sfo, EPTA:2023xxk}, PPTA~\cite{Reardon:2023gzh, Reardon:2023zen, Zic:2023gta} and CPTA~\cite{Xu:2023wog} collaborations has already shown evidence for a GW stochastic background. 
While the nature of the signal source is yet to be determined, in the reduced parameter space that we consider, $\mathcal{P}_\zeta$ --and therefore the SIWG energy density, see Eq.~\eqref{eq: Omega GW}-- of models compatible with large-scale constraints peak on smaller scales. 
Therefore, we focus on GW observatories at interferometer scales, and in particular compute the signal-to-noise ratio of the SIGW background for LISA and ET\footnote{Cosmic Explorer frequency range and sensitivity are similar to those of ET, and therefore we forego making specific numerical computations for this experiment. The same consideration applies to Taiji and TianQin in relation to LISA.}. 
LISA operates in the  mHz frequency band, while ET covers $1 \lesssim f/\text{Hz}\lesssim 10^{4}$.

Owing to our findings of Sec.~\ref{sec:Non-Gaussianities}, we assume that corrections to the spectrum of SIGW due to non-Gaussianity of $\zeta$ are sub-leading~\cite{Adshead:2021hnm, Chang:2023aba, Perna:2024ehx}.
We assume the SIGW are produced at horizon re-entry during radiation domination, and the fraction of SIGW energy density is~\cite{Ananda:2006af, Baumann:2007zm} 
\begin{align}
\label{eq: Omega GW}
     \Omega_{\mathrm{GW}} (k) = c_{g}\,  \Omega_{\mathrm{rad}, 0} \int_{0}^{1} \dd d  \int_{1}^{\infty} \dd s \, \mathcal{P}_\zeta \lp \frac{s - d}{2} k \rp \mathcal{P}_\zeta \lp \frac{s + d}{2} k \rp {\cal{T}}_{\mathrm{rad}}(d, s) \;,
\end{align}
where $c_{g}$ depends on the number of relativistic species at the time of GW production, the radiation energy density today is $\Omega_{\mathrm{rad}, 0} = 8 \times 10^{-5}$ and ${\cal{T}}_{\mathrm{rad}}$ is an appropriate transfer function.  
By assuming the Standard Model particle content and that all particles were relativistic during the production of GW, $c_{g} \approx 0.4$. 
We use the publicly available package \texttt{SIGWfast}~\cite{Witkowski:2022mtg} to numerically compute $\ogw$ for models over the $(\chi_0, \, d)$ parameter space we study.

From a given $\Omega_\text{GW}(k)$ signal, one can then compute the signal-to-noise ratio ($\snr$) for each GW observatory, which is defined as (see e.g. Ref.~\cite{Smith:2019wny})
\begin{align}
    \snr = \sqrt{T \int_{0}^{\infty}  \mathrm{d}f\,  \frac{\ogw^2(f)}{\Sigma_{\Omega}^{2}(f)}} \;. 
\end{align}
Here, $T$ is the observation time and $\Sigma_{\Omega}$ refers to the inverse-noise weighted sensitivity to the spectral density, which depends on the instrument's sensitivity and noise response.
We take the noise for LISA from Ref.~\cite{Smith:2019wny}. For the ET, we follow \cite{Caprini:2024ofd, Branchesi:2023mws}, using the latest ET-D data in the context of the activities of the ET Instrument Science Board (ISB)\footnote{The sensitivity curves for the 10km arms are available at \url{https://apps.et-gw.eu/tds/ql/?c=16492}, where they made use of the \texttt{pygwinc} package \cite{2020ascl.soft07020R}.}. 
The expected time of observation with LISA will be $82\%$ of the total mission duration of $4.5$ years.
For ET, we compute the SNR for a 100\% duty cycle over one year.
 
Computing the SNR of a signal is not enough to determine whether it can be detected or not. 
Indeed, a foreground of astrophysical sources (AFG) might hinder the detection, see e.g. Ref.~\cite{Caprini:2024ofd}. 
The AFG for LISA is given by distinct sources: white-dwarf galactic binaries~\cite{Babak:2021mhe, Karnesis:2021tsh} and unresolvable extra-galactic white-dwarf binaries and stellar-mass black hole binaries~\cite{Pieroni:2020rob, Babak:2023lro, Staelens:2023xjn}; for ET it is due to unresolved compact binary mergers~\cite{Branchesi:2023mws, Bellie:2023jlq}. 
The noise from galactic white-dwarf binaries exceeds the LISA noise for a range of frequencies~\cite{Babak:2021mhe}. 
Therefore, it is added to the effective PSD of the instrument that we use.
For the LISA extra-galactic AFG, we find $\text{SNR}^\text{LISA}_\text{WDB+BBH}= 113.3$, while for the ET we find $\text{SNR}^\text{ET}_\text{CBM}= 17.8$. 
We classify a model as testable if its SNR is greater than that of expected AFGs (which we label $\snrafg$). 

\begin{figure}
\centering
\captionsetup[subfigure]{justification=centering}
\begin{subfigure}[b]{0.49\textwidth}
\includegraphics[width=\textwidth]{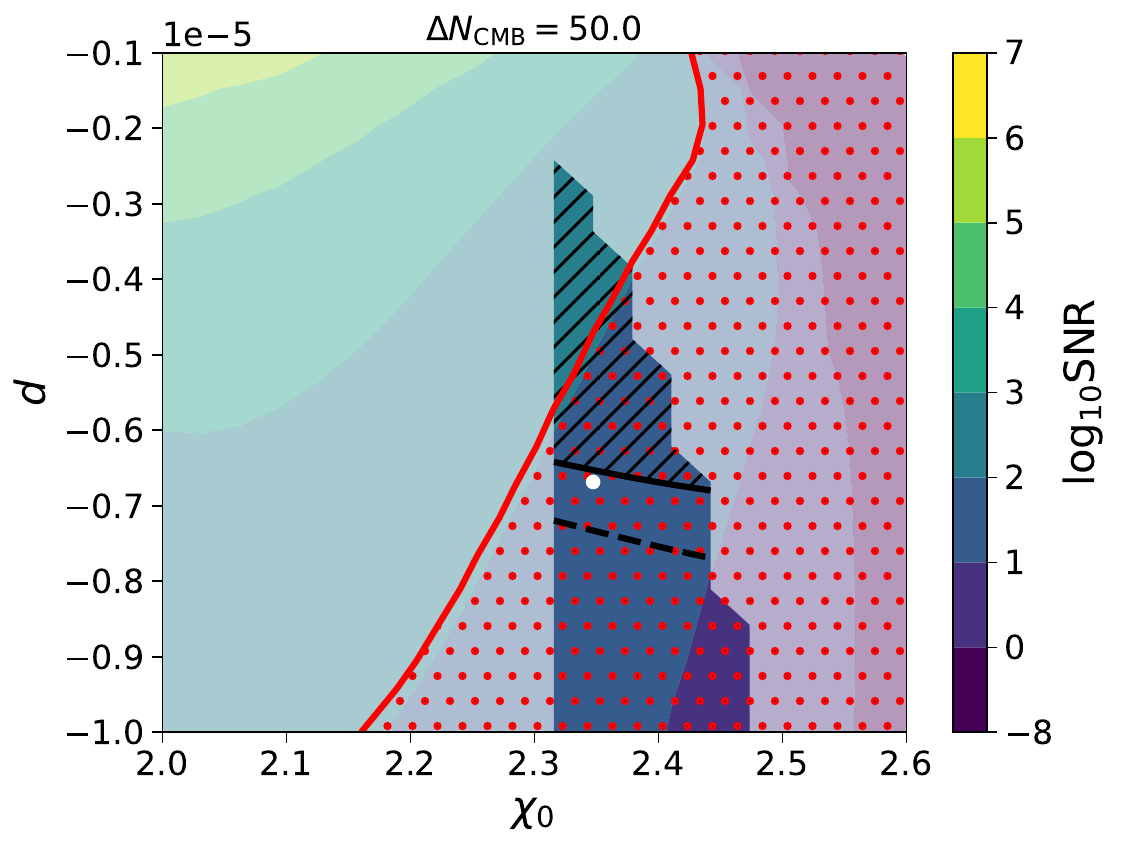}
\end{subfigure}
\begin{subfigure}[b]{0.49\textwidth}
\includegraphics[width=\textwidth]{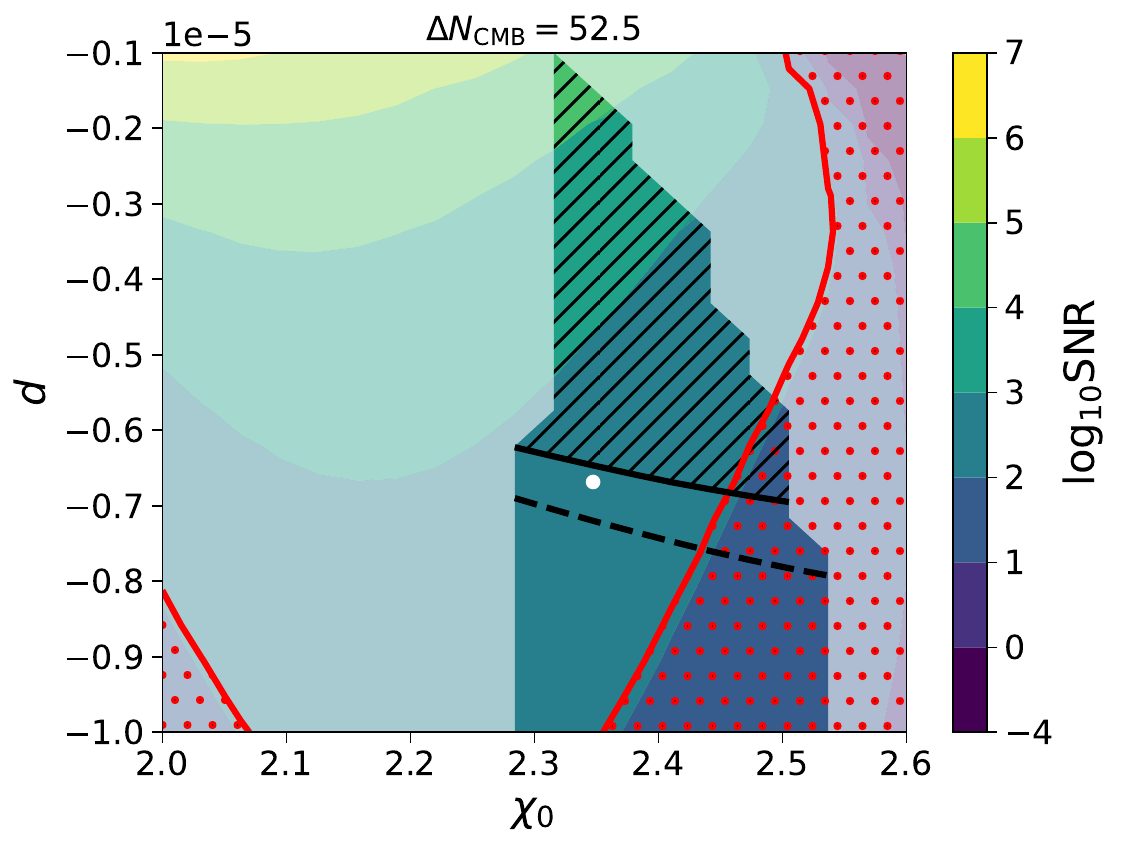}
\end{subfigure}
\begin{subfigure}[b]{0.49\textwidth}
\includegraphics[width=\textwidth]{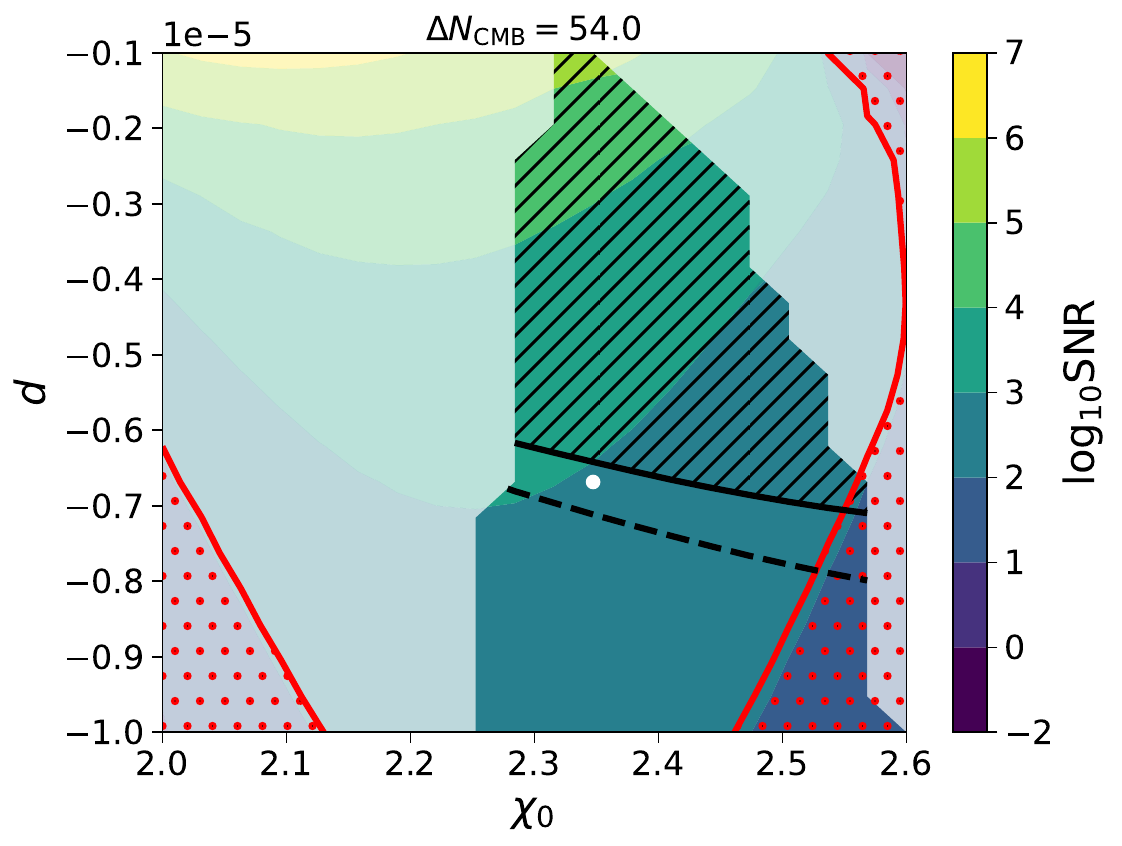}
\end{subfigure}
\caption{Values of $\snr$ for LISA computed over regions of the $(\chi_0,\,d)$ parameter space compatible with the large-scale constraints discussed in Sec.~\ref{sec:large-scale tests}. 
Each panel corresponds to a different value of $\Delta N_\text{CMB}$, see Sec.~\ref{sec: Delta N CMB} for more details on these choices. 
The area hatched with right-angled black dashes is excluded due to PBH overproduction (Sec.~\ref{sec: PBH}), while the continuous and dashed black lines correspond to $f_\text{PBH}=1$ and $f_\text{PBH}=10^{-3}$ respectively. 
The region below the red, solid line (hatched with red dots) corresponds to models for which $\text{SNR}<\snrafg$. 
All models which have no hatching, either black or red, are consistent with the PBH bounds of Sec.~\ref{sec: PBH} and have SNR above that of astrophysical foregrounds. 
In the background of each panel we show with a lighter shade the SNR of models that do not satisfy large-scale constraints.
For clarity of illustration, we identify all models with $\mathrm{SNR}<1$ with a single colour in the colour-bar, regardless of their SNR magnitude. 
}
\label{fig:gw}
\end{figure}
Our results for LISA are represented in Fig. \ref{fig:gw}. 
Once again, each panel corresponds to a different value of  $\Delta N_{\text{CMB}}$. 
First, we note that for a fixed model in the $(\chi_0,\, d)$ parameter space its SNR depends on the duration of reheating, due to the dependence of $k_\text{peak}$ on it (see Fig.~\ref{fig:large scale constraints}). The shorter reheating is, the closer the peak is to LISA scales, and therefore the higher SNR. 
As an example, we select the model with $\{\chi_0 = 2.347,\, d = -6.68 \times 10^{-6}\}$, represented by the white dot in Fig.~\ref{fig:gw}. 
The SNR of the associated signal is $52.19, \, 305.80$ and $874.46$ for $\Delta N_\text{CMB}= 50,\, 52.5$ and $54$ respectively.  
The dependence of $k_\text{peak}$ on $\Delta N_\text{CMB}$ is reflected on the fact that the portion of viable parameter space which LISA will be able to test depends on the duration of reheating. 
No model with $\Delta N_\text{CMB}=50$ and $f_\text{PBH}\leq1$ has SNR sufficiently high, while both models with considerable ($10^{-3}\leq f_\text{PBH}\leq 1$) and negligible PBH production are within reach for LISA when reheating is brief ($\Delta N_\text{CMB}=52.5$ and $54$). 

\begin{figure}
\centering
\captionsetup[subfigure]{justification=centering}
\begin{subfigure}[b]{0.49\textwidth}
\includegraphics[width=\textwidth]{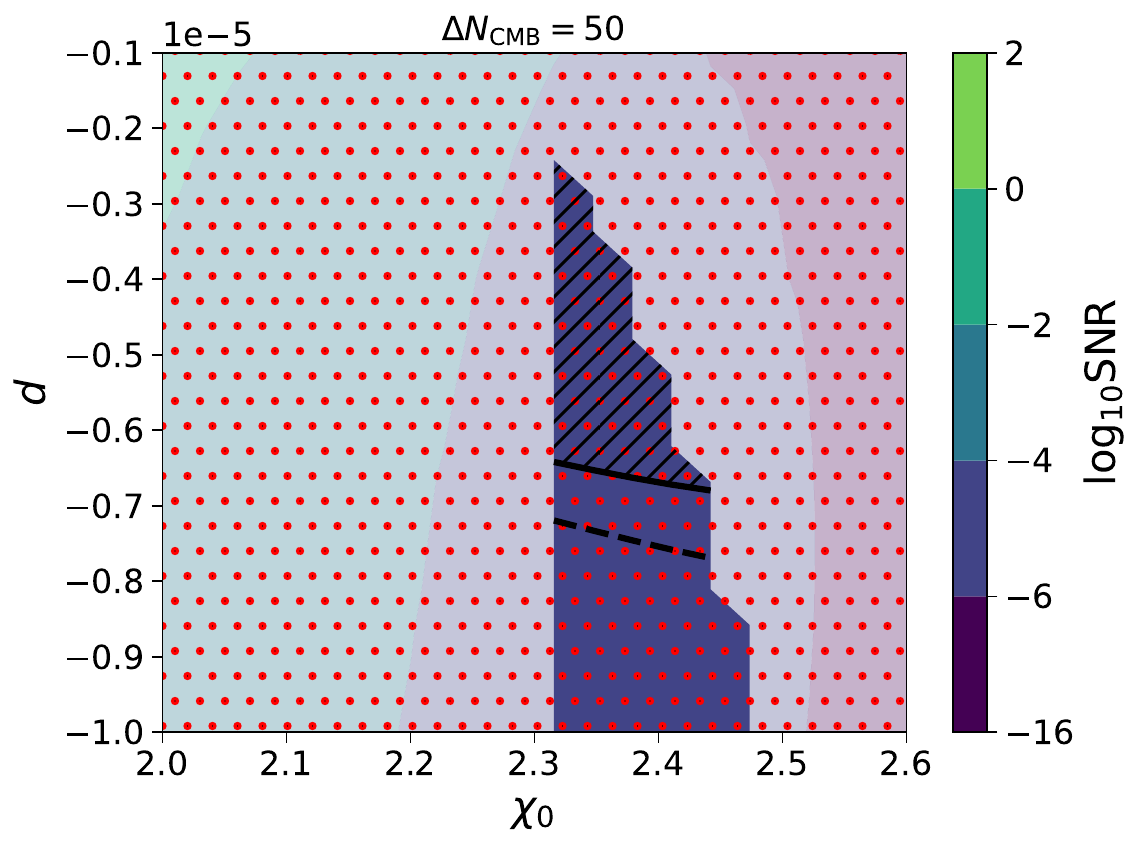}
\end{subfigure}
\begin{subfigure}[b]{0.49\textwidth}
\includegraphics[width=\textwidth]{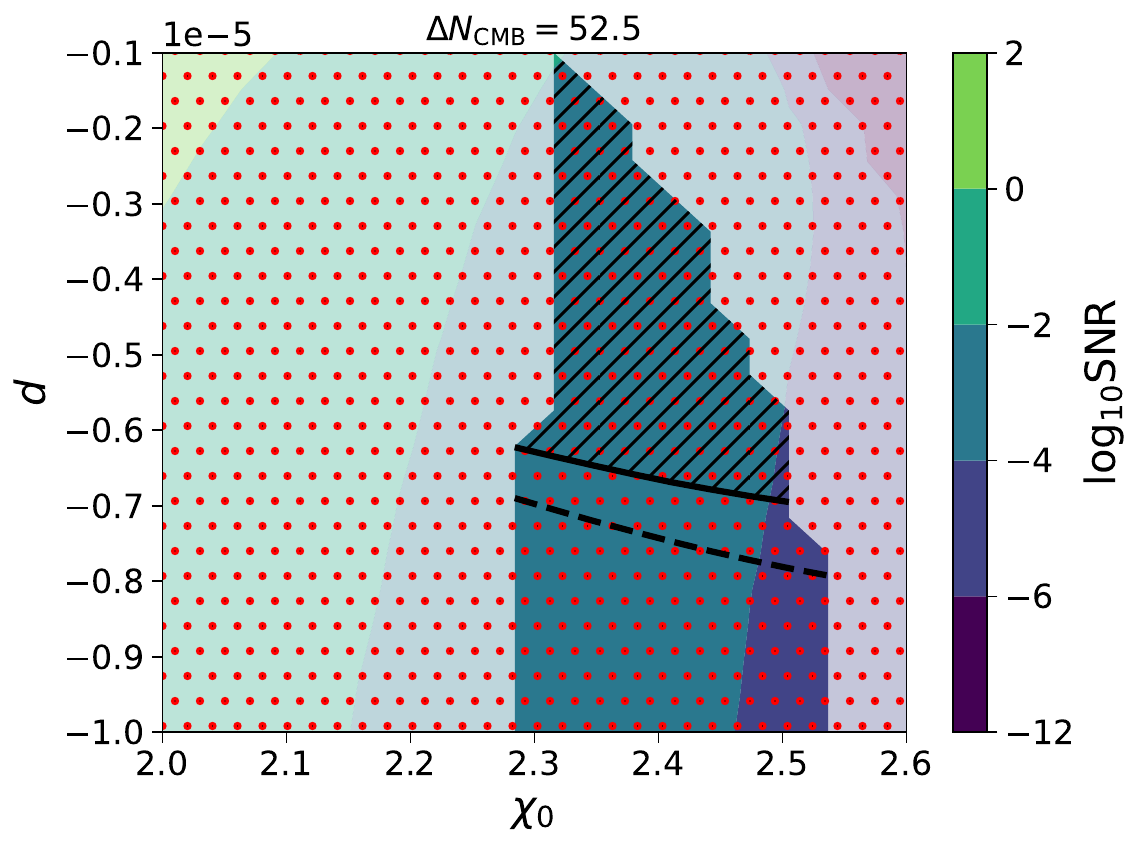}
\end{subfigure}
\begin{subfigure}[b]{0.49\textwidth}
\includegraphics[width=\textwidth]{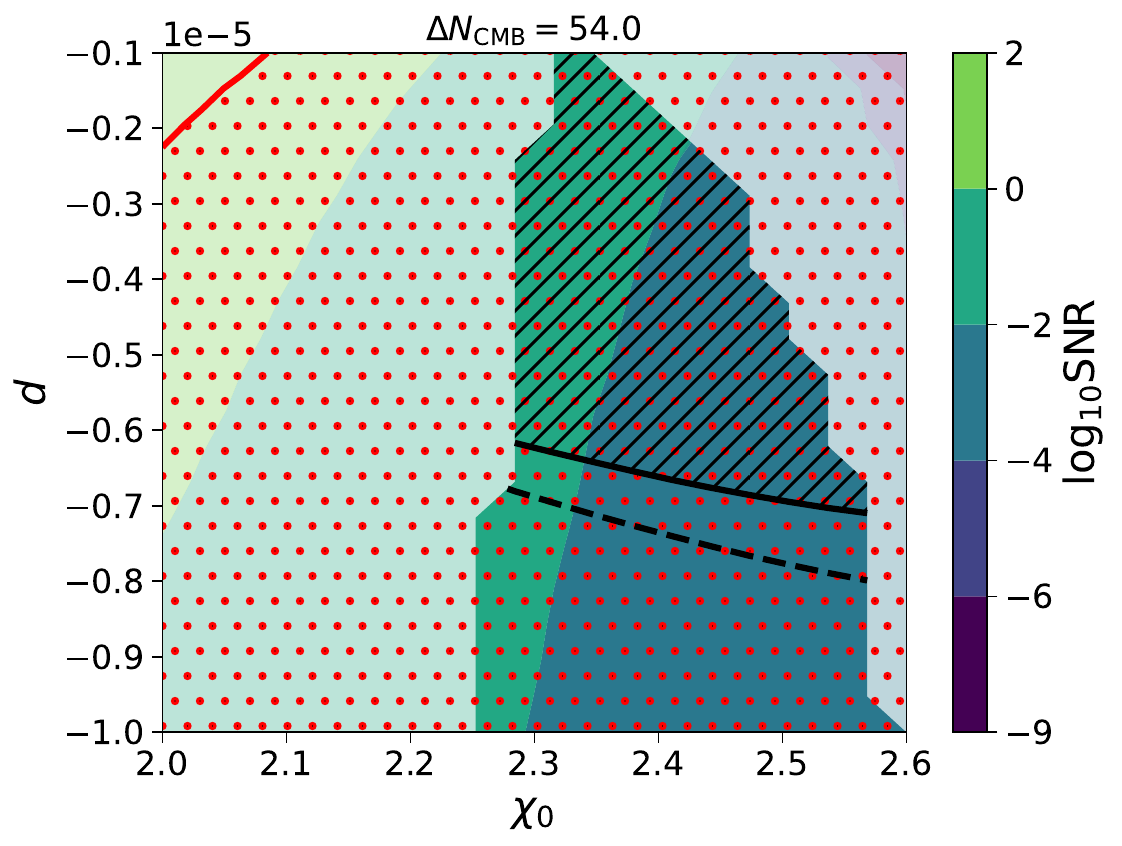}
\end{subfigure}
\caption{Values of $\snr$ for ET computed over regions of the $(\chi_0,\,d)$ parameter space compatible with the large-scale constraints discussed in Sec.~\ref{sec:large-scale tests}. 
The area hatched with right-angled black dashes is excluded due to PBH overproduction (Sec.~\ref{sec: PBH}), the continuous and dashed black lines correspond to $f_\text{PBH}=1$ and $f_\text{PBH}=10^{-3}$ respectively, and the region below the red, solid line (hatched with red dots) corresponds to models for which $\text{SNR}<\snrafg$. 
In the background we show with a lighter shade the SNR for models that do not satisfy large-scale constraints.
For clarity of illustration, we identify all models with $\mathrm{SNR}<10^{-6}$ with a single colour in the colour-bar, regardless of their SNR magnitude. }
\label{fig:ETSNR}
\end{figure}
We performed the same analysis for ET, see Fig.~\ref{fig:ETSNR}. 
For $\Delta N_\text{CMB} = 50$, the SNR over the whole parameter space ranges from $10^{-16}$ to $10^{-2}$. 
Higher SNR is produced for shorter reheating stages, for the same reason explained above for LISA.
For $\Delta N_\text{CMB} = 52.5$ we get $10^{-12} \lesssim \snr \lesssim 10 $, and for $\Delta N_\text{CMB} = 54$ we obtain $10^{-9}\lesssim \snr \lesssim  100$. 
However, no model which is compatible with large-scale constraints also has $\snr > \snrafg$. 
The fact that LISA will be able to constrain the parameter space we consider, while ET will not, is due to the fact that viable models peak around LISA scales (see Fig.~\ref{fig:large scale constraints}), while ET operates at higher frequencies.

\section{Discussion}
\label{sec:Conclusions}
When investigating the small-scale phenomenology of inflationary models, one must simultaneously take into account current observations on all scales, and theoretical consistency requirements. 
In this work we present our approach to this issue (see Sec.~\ref{sec:methodology}), and showcase it by studying the hybrid $\alpha$-attractors introduced in Ref.~\cite{Braglia:2022phb}. 

Our findings can be summarised as follows: 
\begin{itemize}
    \item Current large-scale CMB measurements and upper limits on CMB spectral $\mu$-distortions constrain the parameter space, see Fig.~\ref{fig:large scale constraints}.
    Models with $\chi_0\lesssim 2.3$ are excluded because they predict a scalar spectral tilt which is too red. 
    For larger $\chi_0$, $\mu$-distortions constraints are the strongest, with the constraints from $\alpha_s$ comparable or marginally stronger when $|d|$ is large.  
    Models in this area of the parameter space display multi-field dynamics already at CMB scales, which is why they are inconsistent with current observations. 
    The portion of parameter space compatible with large-scale observations shrinks for longer reheating stages, i.e. smaller $\Delta N_\text{CMB}$.
    The size of non-Gaussianity and amplitude of primordial GWs at CMB scale are in agreement with observations, and they therefore do not yield to further constraints on the parameter space.
    \item At peak scales, we find non-Gaussianity of the local type, with very small amplitude, $f_\text{NL}\sim \mathcal{O}(0.1)$, and mild dependence on $\chi_0$, see Fig.~\ref{fig:fNLChi0D}. 
    Results for $f_\text{NL}$ allow us to establish that non-linear effects due to cubic interactions on small scales are negligible. 
    Importantly, this implies that the tree-level scalar power spectrum can be safely employed to explore the small-scale phenomenology. 
    This is the first study of non-Gaussianity for hybrid $\alpha$-attractors. 
    \item Excluding models which overproduce PBHs further constrains the parameter space, see Fig.~\ref{fig:fPBH}. 
    Regardless of the reheating duration, we find that some of the models viable on large scales can lead to significant PBH production, e.g. $10^{-3}\leq f_\text{PBH}\leq 1$.
    \item  Our results of SNR show that some models featuring a short reheating stage ($\Delta N_\text{CMB}=52.5$ and $54$), and which comply with all the tests described above, are within reach for LISA, see Fig.~\ref{fig:gw}. 
    While a fraction of these also leads to considerable PBH formation, the detection and characterisation of SIGW will enable us to test models that produce a negligible amount of PBHs as well.
    For the slice of parameter space we study, we find that no model compatible with large-scale constraints also leads to a detectable signal for ET, see Fig.~\ref{fig:ETSNR}. 
\end{itemize}

\noindent \textbf{Future directions.} 
There are several interesting extensions of the present work. 

Let us first comment on the methodology. 
In this work, we have illustrated the effect of a matter-dominated reheating stage by considering three different values of $\Delta N_\text{CMB}$. 
These have been carefully selected by ensuring that they are compatible with the parameter space we consider, see Sec.~\ref{sec: Delta N CMB}. 
In future work, we plan to describe reheating following the parametrisation introduced in Ref.~\cite{Martin:2014nya}, and systematically include the reheating parameter alongside other inflationary parameters (see also Refs.~\cite{Iacconi:2023mnw, Ballardini:2024ado}). 
Furthermore, while we have constrained the parameter space by means of marginalised measurements/upper bounds obtained from current data, Bayesian inference for inflation can also be performed (see e.g. Refs.~\cite{Martin:2013nzq,Martin:2024qnn}). 
This would allow us to derive the current posterior distribution for the inflationary parameters, and to perform Bayesian model comparison.
Finally, as commented at the end of Sec.~\ref{sec: PBH}, for parameter-space regions leading to peaks on very small scales, e.g. $k_\text{peak}\gtrsim 10^{16}\,\text{Mpc}^{-1}$, the inclusion of observational limits on the PBH abundance as a function of the PBH mass would potentially yield to further constraints.

In this work we have illustrated the principles of our methodology by applying it to a parameter-space slice of hybrid $\alpha$-attractors. 
In order to perform comprehensive tests of hybrid $\alpha$-attractors, we will apply our methodology to the whole multi-dimensional parameter space.
Results of the present work already demonstrate that areas of the parameter space which are consistent with the tests performed also produce a SIGW signal within reach for LISA, see Fig.~\ref{fig:gw}. 
It would be interesting to assess how the model parameters will be constrained in event of a GW detection (see, e.g., Ref.~\cite{LISACosmologyWorkingGroup:2024hsc}). 
Furthermore, while in this work we have focused on exponential hybrid $\alpha$-attractors, our method can be applied to any model of interest. 
For example, it would be interesting to explore the parameter space of polynomial hybrid attractors~\cite{Braglia:2022phb}, as well as other models leading to enhanced fluctuations on small scales.

\section*{Acknowledgments}
The authors are particularly grateful to Matteo Braglia and Andrew Gow for numerous discussions, and to Tessa Baker, Ian Harry, Isabela de Matos, David Mulryne, David Wands and Lukas T. Witkowski for useful correspondence. We also thank Toby Maule, Xan Morice-Atkinson and Guilherme Brando for sharing their knowledge on the use of High Performance Compute clusters and Python coding.
LI was supported by a Royal Society funded postdoctoral position and acknowledges current financial support from the STFC under grant ST/X000931/1.
LFG was suported by Conselho Nacional de Desenvolvimento Científico e Tecnológico - CNPq and by Fundação de Amparo à Pesquisa e Inovação do Espírito Santo (Fapes) - Edital nº 15/2022. 
FTF acknowledges financial support from the National Scientific and Technological Research Council (CNPq, Brazil).
We acknowledge the use of \texttt{PyTransport}~\cite{Mulryne:2016mzv, Ronayne:2017qzn, Dias:2016rjq} and \texttt{SIGWfast}~\cite{Witkowski:2022mtg}. 
Numerical computations were done on the Sciama High Performance Compute (HPC) cluster, which is supported by the ICG, SEPNet and the University of Portsmouth.

\bibliography{refs} 
\bibliographystyle{JHEP}

\end{document}